%% file: ms.tex
\documentclass[10pt,twocolumn,journal]{IEEEtran}

\usepackage{amsmath, mathrsfs,balance}
\usepackage{bm}

\usepackage{graphicx}
\usepackage{color}
\usepackage{tikz}
\usepackage{pgfplots}
\pgfplotsset{compat=newest}
\usetikzlibrary{patterns}
\usetikzlibrary{positioning}
\usetikzlibrary{datavisualization}
\usetikzlibrary{datavisualization.formats.functions}
\usetikzlibrary{backgrounds}
\usetikzlibrary{shapes,snakes}

\makeatletter
\def\maketag@@@#1{\hbox{\m@th\normalfont\normalsize#1}}
\makeatother

\usepackage[none]{hyphenat}

\newcommand{\subparagraph}{}
\usepackage[compact]{titlesec}
\titlespacing*{\section}{2pt}{1\baselineskip}{0.9\baselineskip}

\allowdisplaybreaks

\usepackage{amsfonts}
\usepackage{amssymb}
\usepackage{color}
\usepackage{multirow}
\usepackage{graphicx}

\usepackage{caption}
\usepackage{subcaption}
\usepackage[numbers,sort&compress]{natbib}
\usepackage{balance}

\input{symbols.tex}

\input{macros.tex}

\def\gUE{g^{\text{UE}}}
\def\gBS{g^{\text{BS}}}

\def\snraftco{{\mathsf{SNR}_\text{ABF}}}
\def\snraftce{{\mathsf{SNR}^\text{CE}_\text{ABF}}}

\def\snrbef{{\mathsf{SNR}_\text{BBF}}}
\def\snrbefco{{\mathsf{SNR}_\text{BBF}}}

\usepackage{tikz}
\usepackage{pgfplots}
\pgfplotsset{compat=newest}
\usetikzlibrary{patterns}
\usetikzlibrary{positioning}
\usetikzlibrary{datavisualization}
\usetikzlibrary{datavisualization.formats.functions}
\usetikzlibrary{backgrounds}
\usetikzlibrary{shapes,snakes}
\usepackage{textcomp}
\usetikzlibrary{automata}

\allowdisplaybreaks

\usepackage{algorithm}
\usepackage{algpseudocode}
\usepackage{pifont}
\usepackage{varwidth}

\def\one{{\bf 1}}

\usepackage{multicol,lipsum}

\def\herm{{\sfH}}

\def\ptot{{P_{\ttt \tto \ttt}}}
\def\pdim{{P_{\ttd \tti \ttm}}}

\def\cg{{\clC\clN}} 
\def\matlab{{MATLAB\textcopyright\,}}

\usepackage{color}

\begin{document}

\title{A Scalable and Statistically Robust Beam Alignment Technique for mm-Wave Systems}
\author{Xiaoshen Song, Saeid Haghighatshoar,  \IEEEmembership{Member, IEEE,} Giuseppe Caire, \IEEEmembership{Fellow, IEEE} 
}

\maketitle
\begin{abstract}
Millimeter-Wave (mm-Wave) frequency bands provide an opportunity for much wider channel bandwidth compared with the traditional  sub-6 GHz band. Communication at mm-Waves is, however,  quite challenging due to the severe propagation path loss. To cope with this problem, directional beamforming both at the \textit{Base Station} (BS) side and at the user side is necessary in order to establish a strong path conveying enough signal power. Finding such beamforming directions is referred to as the \textit{Beam Alignment} (BA) and is known to be a challenging problem. This paper presents a new scheme for efficient BA, 
based on the estimated second order channel statistics. As a result, our proposed algorithm is highly robust to variations of the channel time-dynamics compared with other proposed approaches  based on the estimation of the channel coefficients, rather than of their second-order statistics. 
In the proposed scheme, the BS probes the channel in the \textit{Downlink} (DL) letting each user to estimate its own path direction. All the users within the BS coverage are trained simultaneously, without requiring ``beam refinement'' with multiple interactive rounds of \textit{Downlink}/\textit{Uplink} (DL/UL) transmissions, 
as done in other schemes. Thus, the training overhead of the proposed BA scheme  is independent of the number of users in the system. We pose the channel estimation at the user side as a  \textit{Compressed Sensing} (CS)  of a non-negative signal and use the recently developed \textit{Non-Negative Least Squares} (NNLS) technique to solve it efficiently.  The performance of our proposed algorithm is assessed via computer simulation in a relevant mm-Wave scenario. 
The results illustrate that our approach is superior to the state-of-the-art BA schemes proposed in the literature in terms of 
training overhead in multi-user scenarios and robustness to variations in the channel  dynamics.
\end{abstract}


\section{Introduction}\label{introduction}

Communication at millimeter-waves (mm-Waves for short) 
provides an opportunity to fulfill the demand for high data rates in the next generation communication networks because of the 
large available bandwidth \cite{OverviewHeath}. A critical challenge to signaling at mm-Waves compared with sub-6\,GHz spectrum 
is the severe propagation loss \cite{rappaport2013millimeter}. Fortunately, due to the small wavelength, it is possible to package a large number
of antenna elements in a small form factor. Such large arrays can be used to  provide high-gain directional beamforming at both the \textit{Base Station} (BS) side and the {\em User Equipment} (UE) side in order to boost the \textit{Signal-to-Noise Ratio} (SNR) to sufficiently high levels, such that small-cell outdoor communication is possible.  Moreover, it has been observed experimentally and modeled mathematically that the propagation channel at mm-Waves  is formed by 
a very sparse collection of scatterers in the angle domain \cite{SayeedVirtualBeam2002,Nitsche2014Overview, ChenFourierbasis, SaeidBA2016}. 
This implies that, to establish reliable communication, the BS and the UE need to focus their beams in the direction of a strong path. 
For example, in the case of \textit{Line-of-Sight} (LoS) propagation, the beams must point at each other since the LoS path is typically the strongest one. 
More in general, we refer to the problem of finding a narrow beam direction at both the BS and the user side yielding an SNR {\em after beamforming} 
above a desired threshold as the \textit{Beam Alignment} (BA) problem. This problem is quite well studied in the literature \cite{SayeedVirtualBeam2002,Nitsche2014Overview, SaeidBA2016,Desai2014LimitRF,wang2009beam, chen2011multi, hur2013millimeter, alkhateeb2014channel, KokOverlap,xiarobust2, BPDAMultiuser,  ahmedmultiuser}. In particular, it is known to be a challenging problem since in mm-Waves the SNR 
{\em before beamforming} is typically very low, especially in outdoor non-LoS conditions. 
Moreover, although the number of array antennas may be very large, 
the number of \textit{Radio Frequency} (RF) chains is limited, due to the difficulty of implementing a full RF 
chain (including A/D conversion, modulation, and PA/LNA amplification) for each  array element in a very small form factor and for a very large bandwidth. 
The small number of RF chains prevents the implementation of classical 
digital beamforming schemes in the baseband domain. In contrast, \textit{Hybrid-Digital-Analog} (HDA) beamforming must be 
considered \cite{Desai2014LimitRF,MolischSurvey}.  In particular, a naive sequential scanning of the \textit{Angle-of-Departure} (AoD) and \textit{Angle-of-Arrival} (AoA) domains with narrow beams in order to find an alignment to a strongly connected propagation path is very time consuming and unfeasible in practice. 

\subsection{Related State-of-the-Art}

The inefficiency of naive alignment search has motivated BA algorithms based on hierarchical adaptive search, interactive search, and \textit{Compressed Sensing} (CS) techniques \cite{wang2009beam, chen2011multi, hur2013millimeter, alkhateeb2014channel, KokOverlap,xiarobust2, BPDAMultiuser,  ahmedmultiuser}. 

The fundamental idea of hierarchical methods is to use wider beam patterns at the start of the search and to refine them in several consecutive stages. 
In \cite{alkhateeb2014channel}, for example, the authors develop a bisection algorithm in which the range of AoDs and AoAs are divided by a factor of $2$ at each step and  is refined by probing the resulting $2\times2$ sections and identifying the section with the maximum received power. A similar idea using overlapped beam patterns is used in \cite{KokOverlap}. Such hierarchical techniques, however, require the interaction of the BS with each individual user, since the training is bi-directional and involves both \textit{Downlink} (DL) probing and \textit{Uplink} (UL) feedback for each iterative round. Therefore, it is not obvious how to  extend/adapt these approaches to a multiuser scenario, where a BS has to train 
potentially many users at a time. 

In \cite{xiarobust2}, a method is proposed where the BS and the UE iteratively and collaboratively 
identify the dominant eigenvector of their channel matrix via the well-known {\em power method}. However, this approach requires to
demodulate the signal at each antenna both at the BS and at the UE side. 
Therefore, this method is essentially incompatible with the HDA beamforming architecture.  

More recently, considering the natural channel sparsity in the AoA-AoD domain \cite{SayeedVirtualBeam2002,Nitsche2014Overview, ChenFourierbasis, SaeidBA2016}, CS-based algorithms have been proposed for BA in mm-Waves \cite{BPDAMultiuser, ahmedmultiuser, ChoiBA2015, HeathPhaseSwitch2016, HeathWidebandBA2017}. These algorithms are efficient and particularly attractive for multi-user scenarios, 
but they are based on the assumption that the instantaneous channel remains invariant during the whole probing/measurement stage  (the same assumption is also adopted in \cite{alkhateeb2014channel,KokOverlap}). This assumption is typically not satisfied in practice 
due to the large Doppler spread at mm-Waves, implying fast time-variations of the channel coefficients \cite{WeilerMeasure2014, HeathVariation2017}. 
It should be noticed here that the channel time-variations are greatly reduced {\em after BA is achieved}, since once the beams are aligned, 
the effective channel angular spread is very small \cite{HeathVariation2017} (e.g., in the case of LoS propagation the Doppler reduces to a simple frequency shift which can be  estimated and compensated by standard carrier synchronization schemes). 
Nevertheless, {\em before BA is achieved} the channel variability over time can be large, since even a small motion of a few centimeters traverses 
several wavelengths, potentially producing multiple deep fades \cite{WeilerMeasure2014}. Moreover, a naive application of the conventional CS techniques typically results in a wide spread of the transmitted signal power during the probing stage in the angle domain and diminishes the SNR of the resulting measurements. 
This is not problematic in sub-6\,GHz but it may be a big problem in mm-Waves due to the very low SNR {\em before beamforming}. 
Interestingly, this low-SNR problem is widely overlooked in the CS-based channel estimation literature, which typically assumes unrealistic values of the 
pre-beamforming SNR. 

\subsection{Contributions}

In this paper, we propose a novel BA scheme that has the following advantages compared with the existing works in the literature:

%
%

{1) \textit{System-level Scalability:} In our approach, during the BA phase, the BS actively probes 
the channel by periodically broadcasting a {\em pseudo-random beamforming codebook}, i.e., a sequence of pseudo-random beam 
patterns, over reserved {\em beacon} slots in the DL, while all users stay in listening mode. In particular, all users can collect
a sufficient number of measurements in order to estimate their own relevant channel information, namely, the AoA-AoD of a strong scatterer 
conveying sufficient signal power from/to the BS. Since there is no need for interaction between the BS and each user, the proposed 
BA scheme is highly scalable and its overhead and complexity do not grow with the number of active users in the system.

{2) \textit{User-specific beamforming codebook:} During the beacon slots, each user can apply its own {\em receive beamforming codebook}, 
given again by a sequence of random beam patterns. In the proposed scheme, 
each user selects its receive beamforming codebook based on the number of available RF chains and on its SNR. 
In brief, when a user is close to the BS and has a sufficiently high pre-beamforming SNR, it can use wider beams  in order to speed up the channel 
estimation by taking less measurement rounds in time. 
In contrast, when a user is far from the BS and has a very low pre-beamforming SNR, it applies narrower receive beams to obtain measurements 
with a higher beamforming  gain and thus achieving sufficiently large SNR. In particular, we shall see that for a specific SNR level before beamforming, 
there is an optimal beam spreading factor that results in the fastest channel acquisition. On the practical side, our method has the advantage that the
receiver beamforming strategy at the users can be individually tailored to each specific user, depending on its hardware (how many RF chains)
and on its pre-beamforming SNR conditions, without impacting the overall system functions. 

{3) \textit{Robustness to Variations in Channel Statistics:} As explained before, most of the existing works in the literature use the assumption that the instantaneous channel coefficients remain invariant during the whole BA phase. This is difficult to meet in mm-Waves due to the large carrier frequency, large Doppler spread, and 
consequently fast channel variations \cite{WeilerMeasure2014, HeathVariation2017}. Our scheme makes use of the channel 
second order statistics and is highly robust to variations in channel time-dynamics. We also illustrate via numerical simulations that  CS-based algorithms  fail to estimate the channel strong path direction when the channel is significantly time-varying, i.e., it undergoes several fading cycles during the estimation period, 
whereas our scheme performs well for a wide range of channel dynamics. 

{4) \textit{Low-complexity Channel Estimation and Protocol Simplicity:} 
In our scheme, each user needs to estimate the channel from its received measurements, thus, all the computation is done at the user side. We show that the resulting channel estimation boils down to a \textit{Non-Negative Least Squares} (NNLS) problem, which can be solved very efficiently via standard techniques in the literature. After estimating its best beam direction, each user communicates the corresponding beam index 
through a random access UL slot  (see Section \ref{Mathematical}), which can already benefit from full beamforming gain 
at the user side and from some (limited) beamforming gain at the BS side. At this point, communication can take place with full beamforming 
gain at both sides on regular data slots. 
	
{\bf Notation} We  denote vectors by boldface small (e.g., $\bfa$) and matrices by boldface capital (e.g., $\bfA$) letters.  Scalars are denoted by non-boldface letters (e.g., $a$, $A$). We represent sets by calligraphic letter $\clA$ and their cardinality with $|\clA|$. We denote the empty set by $\emptyset$. We use $\bE$ for the expectation, $\otimes$ for the  Kronecker product of two matrices, $\bfA^\transp$ for transpose, $\bfA^*$ for conjugate, and $\bfA^\herm$ for conjugate transpose 
of a matrix $\bfA$.  The output of an optimization problem such as $\text{arg\,\,min}_{x\in\clX}f(x)$ is denoted by $x^*$.  The complex circularly symmetric Gaussian distribution with a mean $\mu$ and a variance $\gamma$ is denoted by $\cg(\mu,\gamma)$. For an integer $k\in\intgr$, we use the shorthand notation $[k]$ for the set of non-negative integers $\{1,...,k\}$.

\section{Basic Setup}\label{Basic Setup}

\subsection{Channel Model}

We consider a mm-Wave system including a BS equipped with a {\em Uniform Linear Array} (ULA) 
with $M$ antennas and $m$ RF chains where typically $m\ll M$. 
We consider a generic UE, also equipped with an ULA with $N$ antennas and $n\ll N$ RF chains. 
We assume that both the BS and UE arrays have the antenna spacing $d=\frac{\lambda}{2}$, 
where $\lambda$ is the wavelength given by $\lambda=\frac{c_0}{f_0}$, where $c_0$ is the speed of the light and where $f_0$ is the carrier frequency. 
We denote by $\theta, \phi \in [-\frac{\pi}{2}, \frac{\pi}{2}]$ the steering angles with respect to the BS and UE arrays. We represent the array response of the BS and UE arrays to a planar wave coming from the angles $\theta$ and $\phi$ with respect to the BS and UE with the $M$-dim and $N$-dim array vectors 
$\bfa(\theta) \in \bC^{M}$ and $\bfb(\phi) \in \bC^{N}$ respectively, with elements
\begin{align}
[\bfa(\theta)]_k&=e^{j (k-1)\pi \sin(\theta)}, k \in[M], \label{a_resp_BS}\\
[\bfb(\phi)]_l&=e^{j (l-1)\pi \sin(\phi)},\  l\in[N].\label{a_resp_UE}
\end{align}
We assume that the communication between the BS and the UE occurs via a collection of sparse \textit{multi-path components} (MPCs) in the AoA-AoD and delay domain \cite{OverviewHeath}, where the $N \times M$ low-pass equivalent impulse response of the channel at a symbol time $s$ is given by
\begin{align}\label{ch_mod_disc_mp}
\sfH_s(\tau)&=\sum_{l=1}^L \rho_{s,l} \bfb(\phi_l) \bfa(\theta_l)^\herm \delta(\tau-\tau_l),
\end{align}
where $\rho_{s,l}$ is the random channel gain of the $l$-th MPC at AoA-AoD-delay $(\theta_l, \phi_l, \tau_l)$, $l \in [L]$. 
Typically the number of {\em significant} MPCs satisfies $L \ll \max\{M,N\}$ \cite{rappaport2013millimeter}. 
In practice there may be a large number of MPCs that convey such a small amount of signal power that can be simply neglected since in any case
they will not be useful for signal transmission even after the BA is achieved.  Note that in \eqref{a_resp_BS}, \eqref{a_resp_UE}, and \eqref{ch_mod_disc_mp} 
we made the implicit assumption (very common in most beamforming and array processing literature) that the communication bandwidth denoted by $B$ is much smaller than the carrier frequency $f_0$ (i.e., $B \ll f_0$) such that the array response is essentially frequency-invariant, i.e., the relative change of the wavelength $\lambda$ over the frequency interval $f \in [f_0 - B/2, f_0 + B/2]$ is negligible. We adopt a block fading model, where the channel gains $\rho_{s,l}$, $l \in [L]$, remain invariant over the \textit{coherence time} of the channel of duration $\Delta t_c$ but change randomly across different \textit{coherence times} according to a given 
wide-sense stationary process with given Doppler power spectral density \cite{molisch2012wireless}. 
We also assume that each MPC is formed by a cluster of a large number of micro-scatterers corresponding (roughly) to the same delay and AoA/AoD,  such that the channel gains ${\rho_{s,l} \sim \cg(0, \gamma_l)}$ have a zero-mean complex Gaussian distribution. 
The channel model \eqref{ch_mod_disc_mp} can be extended to the case where there is a continuum of MPCs connecting 
the BS and the UE, where the channel model is given by   
\begin{align}\label{ch_mod_contc_mp}
\sfH_s(d \tau)=\rho_{s}(d\theta,d\phi, d\tau)  \bfb(\phi) \bfa(\theta)^\herm,
\end{align}
where $\rho_{s}(d\theta,d\phi, d\tau)$ denotes the angle-delay random impulse response of the channel.
The BA algorithm developed in this paper holds for the general case \eqref{ch_mod_contc_mp} as long as the angular scattering function
(i.e., the channel power distribution) has a small support over the AoA-AoD domain. 
The small support of the channel angular scattering function is motivated by experimental observations, suggesting that the propagation at mm-waves 
occurs along MPCs with small AoA-AoD spread, as illustrated in Fig.~\ref{scheme}.
For the sake of simplicity of exposition,  we will use the discrete MPC model (\ref{ch_mod_disc_mp}) in the sequel. 

We also assume that the {\em angle coherence time}, i.e., the time scale over which the 
AoA-AoDs of the scatterers $\{(\theta_l,\phi_l)\}_{l=1}^L$ change significantly, is much longer that the channel coherence time $\Delta t_c$. Hence, 
the angles can be treated as locally constant (but unknown) during the BA phase. 
This {\em local stationarity} of the scattering geometry is widely used in the literature and confirmed 
by channel sounding measurements (e.g., see \cite{HeathVariation2017,DBLP:journals/corr/ShenDSWH17}).

\begin{figure}[t]
	\centerline{\includegraphics[width=7cm]{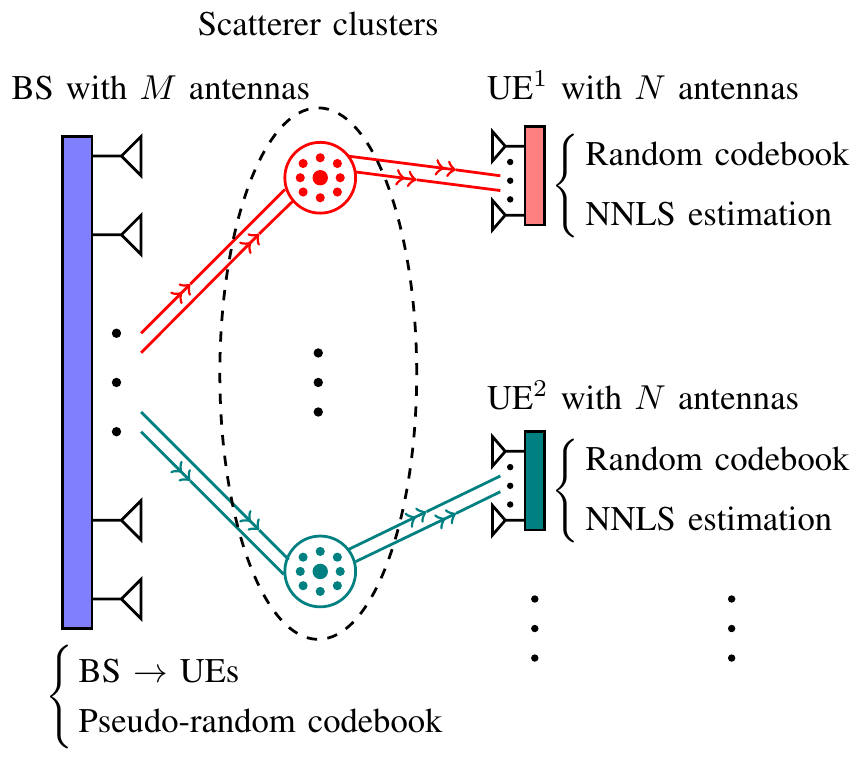}}
	\caption{{\small {\em Illustration of the physical channel model and our proposed \textit{Beam Alignment} (BA) scheme}.}}
\label{scheme}
\end{figure}

\subsection{Signaling Model}

Consider the communication between the BS and a generic UE. 
Since the BS has $m$ RF chains, it can transmit up to $m$ different data streams. For a given signaling interval $t_0$, 
let $x_{s,i}(t)$, $t \in [st_0, (s+1)t_0)$, be the continuous-time baseband equivalent signal corresponding to the $i$-th data stream. 
We assume that the channel is time-invariant over each symbol, i.e., $t_0 < \Delta t_c$. 
To transmit the $i$-th data stream, the BS applies a beamforming vector $\bfu_{s,i} \in \bC^M$.
Without loss of generality, the beamforming vectors are normalized such that $\|\bfu_{s,i}\|=1$.\footnote{Also, note that here we are assuming that the beamforming vectors $\bfu_{s,i}$, $i \in  [m]$ are implemented in the RF domain via an analog beamforming network and therefore they are frequency flat, i.e., they are
constant over the whole signal bandwidth.}
The transmitted signal at symbol time $s$ is  given by 
\begin{align}
\bfx_s(t) & =\sum_{i=1}^m x_{s,i}(t) \bfu_{s,i}.
\end{align}
The received baseband equivalent signal at the UE array is
\begin{align}
\bfr_s(t)&=\int \sfH_s(d\tau) \bfx_s(t-\tau) \nonumber \\
&=\sum_{l=1}^L \rho_{s,l} \bfb(\phi_l) \bfa(\theta_l)^\herm \bfx_s(t-\tau_l) \nonumber \\
&=\sum_{l=1}^L \sum_{i=1}^m \rho_{s,l} x_{s,i}(t-\tau_l) \bfb(\phi_l) \bfa(\theta_l)^\herm \bfu_{s,i} \nonumber \\
&=\sum_{l=1}^L \sum_{i=1}^m \rho_{s,l} \gBS_{s,l,i}  x_{s,i}(t-\tau_l) \bfb(\phi_l) 
\end{align}
where $\gBS_{s,l,i}:=\bfa(\theta_l)^\herm \bfu_{s,i}$ denotes the beamforming gain along the $l$-th MPC  at the BS side for the $i$-th RF chain. 
As stated before, we assume that the UE is also equipped with $n$ RF chains. 
The analog RF signal received at the UE antenna array is distributed to the $n$ RF chain for demodulation. 
This is achieved by signal splitters that divide the signal power by a factor of $n$. 
The noise in the receiver is mainly introduced by the RF chain electronics (filter, mixer, and A/D conversion). It follows that
the noisy received signal at the output of the $j$-th RF chain at the UE side is given by
\begin{align}\label{eq:j_out}
y_{s,j}(t)&=\frac{1}{\sqrt{n}} \bfv_{s,j}^\herm \bfr_s(t)+z_{s,j}(t)\nonumber\\
&= \frac{1}{\sqrt{n}} \sum_{l=1}^L \sum_{i=1}^m \rho_{s,l} \gBS_{s,l,i}  x_{s,i}(t-\tau_l) \bfv_{s,j} ^\herm \bfb(\phi_l) + z_{s,j}(t)\nonumber \\
&=\sum_{i=1}^m \frac{1}{\sqrt{n}} \sum_{l=1}^L \rho_{s,l}   \gBS_{s,l,i} \gUE_{s,l,j} x_{s,i}(t-\tau_l) + z_{s,j}(t) \nonumber \\
&= \sum_{i=1}^m \frac{1}{\sqrt{n}} r_{s,i,j}(t) + z_{s,j}(t)
\end{align}
where ${\bfv_{s,j} \in \bC^N}$ denotes the normalized beamforming vector of the $j$-th RF chain at the UE side, 
where ${\gUE_{s,l,j} :=\bfv_{s,j}^\herm \bfb(\phi_l)}$ denotes the array gain of the  $j$-th RF chain along the $l$-th MPC, where 
${r_{s,i,j}(t):=\sum_{l=1}^L \rho_{s,l} \gBS_{s,l,i} \gUE_{s,l,j}  x_{s,i}(t-\tau_l)}$ denotes 
the signal contribution relative to the $i$-th transmitted data stream of the BS received at the output of the $j$-th RF chain of the UE, 
and where $z_{s,j}(t)$ is the continuous-time complex \textit{Additive White Gaussian Noise} (AWGN) at the output of the $j$-th RF chain,  
with \textit{Power Spectral Density} (PSD) of $N_0$ Watt/Hz. The factor $1/\sqrt{n}$ in (\ref{eq:j_out}) takes into account the power split
said above.

In this paper we consider OFDM signaling with given subcarrier separation $\Delta f$.  
Each symbol $s$ in the general model defined before corresponds here to an OFDM symbol.
The number of subcarriers is given by $F := B / \Delta f$, where $B$ denotes the channel bandwidth as defined before.
We make the standard assumption that the duration $\tau_{\rm cp}$ of the \textit{Cyclic Prefix} (CP) of the OFDM modulation is longer than the channel 
delay spread, implying $t_0 = 1/\Delta f + \tau_{\rm cp}$ with $\tau_{\rm cp} \geq  \max \{\tau_l\} - \min \{\tau_l\}$. 
Hence, after OFDM demodulation, the {\em Inter-Block Interference} is completely removed and we can focus on a per-symbol 
model in the frequency domain \cite{molisch2012wireless}.  
Applying the Fourier transform to the matrix-valued channel impulse response (\ref{ch_mod_disc_mp}), the
frequency-domain  channel matrix at symbol interval $s$ is given by
\begin{align}\label{freq_ch_mod}
\check{\sfH}_s(f)=\sum_{l=1}^L \rho_{s,l} \bfb(\phi_l)\bfa(\theta_l)^\herm e^{-j 2\pi f \tau_l}.
\end{align}
We denote the OFDM subcarriers as $\{f_\omega = \frac{\omega}{t_0} : \omega \in [F]\}$. The channel matrix at subcarrier $\omega$ is given by 
$\bfH_{s}[\omega]:=\check{\sfH}_s(f_\omega)$. 
Let $\check{x}_{s,i}[\omega]$ denote the frequency-domain data symbol for the $i$-th stream. 
Applying OFDM demodulation to the received signal (\ref{eq:j_out}), we obtain
the corresponding  frequency-domain received signal at the $j$-th receiver RF chain, with transmit beamforming vector $\uv_{s,i}$ 
and receive beamforming vector $\vv_{s,j}$ in the form
\begin{eqnarray}\label{receiveblock}
\check{y}_{s,i,j}[\omega] & = & \frac{1}{\sqrt{n}} \bfv_{s,j}^\herm \bfH_{s}[\omega]\bfu_{s,i} \check{x}_{s,i}[\omega]+\check{z}_{s,j}[\omega] \nonumber \\
& = & \frac{1}{\sqrt{n}} \sum_{l=1}^L \rho_{s,l} e^{-j 2\pi \frac{\omega}{t_0} \tau_l} \gBS_{s,l,i} \gUE_{s,l,j} \check{x}_{s,i}[\omega]+\check{z}_{s,j}[\omega], 
\end{eqnarray}
where $\check{z}_{s,j}[\omega] \sim \cg(0, \sigma^2)$ denotes the noise at $j$-th RF chain of UE at subcarrier $\omega$, 
with variance $\sigma^2 = \Delta f N_0$.

\subsection{Beam Alignment}

During the DL probing slots (see  frame structure discussed in Section \ref{Mathematical}), we 
assume that the signal corresponding to different data streams $x_{s,i}(t)$ are orthogonal, i.e., 
\begin{align}
\inp{x_{s,i}}{x_{s,i'}}:=\int_{st_0}^{(s+1)t_0} x_{s,i}(t)^* x_{s,i'}(t) dt=E_{i} \delta_{i,i'},
\end{align}
where $E_{i}$ is the energy per symbol for the $i$-th data stream and $\delta_{i,i'}$ is the Kronecker delta symbol (equal to $1$ for $ i= i'$ and 0 otherwise). 
For example, this can be obtained in the frequency domain by using OFDM and mapping the different streams onto sets of non-overlapping subcarriers. 
We define the SNR \textit{after beamforming} (ABF) for the $i$-th data stream received at the $j$-th RF chain at the UE by 
\begin{align}
\snraftco^{i,j}&:=\frac{\frac{1}{t_0} \bE \left [\int_{st_0}^{(s+1)t_0}  |r_{s,i,j}(t)|^2 dt \right ]}{n N_0 B_i }\nonumber\\
&=\frac{\frac{E_{i}}{t_0}\sum_{l=1}^L \gamma_l |\gUE_{s,l,j}|^2 |\gBS_{s,l,i}|^2}{n N_0 B_i}\nonumber\\
&=\frac{P_{i} \sum_{l=1}^L \gamma_l |\gUE_{s,l,j}|^2 |\gBS_{s,l,i}|^2}{n N_0 B_i },\label{snraft_co}
\end{align}
where $B_i$ and $P_{i}=E_i/t_0$ denote the bandwidth and the average power of $x_{s,i}(t)$, respectively. 
We have $\sum_{i=1}^m P_{i}=\ptot$, where $\ptot$ is the overall transmit power of the BS.
In particular, for equal power allocation ($P_{i}=\ptot/m$) over the streams, we have
\begin{align}
\snraftco^{i,j}&=\frac{\ptot \sum_{l=1}^L \gamma_l |\gUE_{s,l,j}|^2 |\gBS_{s,l,i}|^2}{m n N_0 B_i },\label{snraft_co-equal}
\end{align}
For later use, we also define the SNR before beamforming (BBF) by
\begin{align}\label{snrbef_co}
\snrbefco:=\frac{\ptot \sum_{l=1}^L \gamma_l}{N_0 B}.
\end{align}
This is the SNR obtained when a single data stream ($m = 1$)  is transmitted through a single BS antenna and 
is received in a single UE antenna  (isotropic transmission) over a single RF chain $(n = 1$) with full-band spreading. 

A challenge in mm-Wave communication is that the SNR before beamforming $\snrbefco$ in \eqref{snrbef_co} is typically very low.
This cannot be increased by simply boosting the transmit power $\ptot$ because of hardware and regulation limitations, also because, in general, we would like to design energy-efficient systems. An option to increase the SNR consists of communicating over a smaller bandwidth $B'< B$. However it is well-known that this strategy is suboptimal.\footnote{This statement holds only in the case where the channel coefficients change sufficiently slowly in time. More in general, for time-varying wideband fading channels, it has been shown (e.g., see \cite{medard2002bandwidth,lozano2012non,gomez2017unified,durisi2010noncoherent}) that spreading the transmit power over the entire bandwidth is suboptimal and drives the achievable rate to zero for $B \rightarrow \infty$. Intuitively, this is due to the inability of the receiver to estimate the fading channel coefficients, as explained in \cite{medard2002bandwidth}. The issue of optimal signaling in the presence of time-varying fading is quite intricate and goes beyond the scope of this paper. As a matter of fact, when a large beamforming gain is available at both the BS and the UE side, the effective channel coefficients {\em after beam alignment} are slowly varying (see \cite{HeathVariation2017}) and the SNR after beamforming is large enough, such that the channel can be treated as a standard block-fading AWGN channel known channel coefficients} In fact, assuming a Gaussian channel with SNR equal to $\snrbefco$, 
Shannon's capacity formula yields that the achievable rate in bit/s when communicating over a bandwidth $B'$ is  given by $R = B' \log( 1  + (B/B') \snrbefco)$, which is increasing for $0 < B' \leq B$. Hence, by using a bandwidth smaller than the available channel bandwidth $B$, the achievable rate is reduced. It follows that the only viable alternative consists of using antenna arrays with a large number of antennas both at the BS and at the UE. The goal of BA is to find  \textit{good} beamforming vectors $\bfu_s$ and $\bfv_s$ at the BS and the UE, respectively, in order to boost the SNR by a factor $\approx M$ at the BS side and a factor $\approx N$ at the UE side. This is achieved by aligning the beamforming vectors along the AoA-AoD of a strong MPC of the channel.

\subsection{Sparse Beamspace Representation}\label{Quantization}
\begin{figure}[t]
	\centerline{\includegraphics[width=7.9cm]{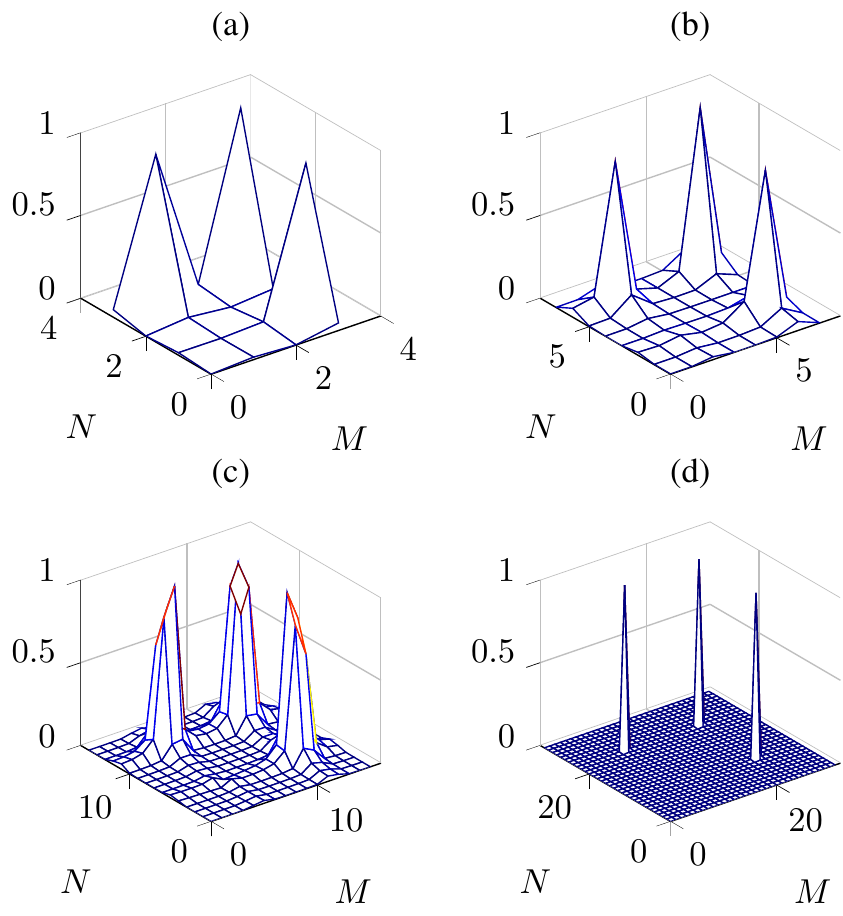}}
	\vspace{-0.15cm}
	\caption{{\small {\em Illustration of the sparsity of the channel matrix $\check{\bfH}_{s}[\omega]$ at an arbitrary subcarrier $\omega$ consisting of $3$ off-grid AoA-AoDs with increasing number of antennas for  $M=N=4$ (a),  $M=N=8$ (b),  $M=N=16$ (c), $M=N=32$ (d)}}.}
	\label{sparsity}
\end{figure}
The AoA-AoDs $(\theta_l,\phi_l)$  in \eqref{freq_ch_mod} take continuous values. 
For later applications in the paper, we need a finite-dimensional representation of the channel. 
Following the well-known approach of \cite{SayeedVirtualBeam2002,bajwa2010compressed,Heath2016overview}, 
known as {\em beamspace representation}, we  obtain such a representation by quantizing the matrix-valued channel response \eqref{freq_ch_mod} with respect to a discrete dictionary in the  AoA-AoD domain. We consider the discrete set of AoA-AoDs 
\begin{align}\label{gridtheta}
\Theta&:=\{\check{\theta}: (1+\sin(\check{\theta}))/2=\frac{k-1}{M}, k\in [M]\},\\
\Phi&:=\{\check{\phi}: (1+\sin(\check{\phi}))/2=\frac{k{'}-1}{N}, k{'} \in [N]\},
\end{align}
and use the corresponding array responses $\clA:=\{\bfa(\check{\theta}): \check{\theta} \in \Theta\}$ and $\clB:=\{\bfb(\check{\phi}): \check{\phi} \in \Phi\}$ as a discrete dictionary to represent the channel response. 
For the ULAs considered in this paper, the dictionary $\clA$ and $\clB$, after suitable normalization, yield orthonormal bases corresponding 
to the columns of the $M\times M$ and $N\times N$  DFT matrices $\bfF_M$ and $\bfF_N$ \cite{ChenFourierbasis}, where 
\begin{align}
[\bfF_{M}]_{k,k'}=\frac{1}{\sqrt{M}}e^{j2\pi (k-1)(\frac{k'-1}{M}-\frac{1}{2})}, k,k'\in[M],\\
[\bfF_{N}]_{k,k'}=\frac{1}{\sqrt{N}}e^{j2\pi (k-1)(\frac{k'-1}{N}-\frac{1}{2})}, k,k'\in[N].
\end{align}
Hence, we obtain the exact representation of the channel matrix $\bfH_{s}[\omega]=\bfF_{N}\check{\bfH}_{s}[\omega]\bfF_{M}^\herm$,  where
\begin{align}\label{htilde}
\check{\bfH}_{s}[\omega]
&=\sum_{l=1}^L \rho_{s,l} e^{-j 2\pi \frac{\omega}{t_0} \tau_l}\check{\bfb}(\phi_l) \check{\bfa}(\theta_l)^\herm,
\end{align}
and where $\check{\bfa}(\theta_l):=\bfF_M^\herm\bfa(\theta_l)$, and $\check{\bfb}(\phi_l):=\bfF_N^\herm\bfb(\phi_l)$ denote the coefficient vectors of the 
array responses $\bfa(\theta_l)$ and $\bfb(\phi_l)$ with respect to the DFT bases, respectively. 
The $m'$-th entry of $\check{\bfa}(\theta_l)$ is given by
\begin{align}
[\,\check{\bfa}(\theta_l)\,]_{m'} &= \frac{1}{\sqrt{M}}\sum_{i=0}^{M-1}e^{-j2\pi i(\frac{m'-1}{M}-\frac{1}{2})}e^{j\pi i\sin(\theta_l)} \nonumber \\
&=\frac{1}{\sqrt{M}}\frac{e^{j\pi\psi_l M}-e^{-j\pi\psi_l M}}{e^{j\pi\psi_l }-e^{-j\pi\psi_l }}e^{-j\pi\psi_l (M-1)} \nonumber \\
&=\frac{1}{\sqrt{M}}\frac{\sin(\pi\psi_l M)}{\sin(\pi\psi_l )}e^{-j\pi\psi_l (M-1)},\label{loc_kernel}
\end{align}
where ${\psi_l =\frac{m'-1}{M}-\frac{1}{2}\sin(\theta_l)-\frac{1}{2}}$. A similar expression holds for $\check{\bfb}(\phi_l)$. It is seen from \eqref{loc_kernel} that 
$|[\,\check{\bfa}(\theta_l)\,]_{m'}|=\frac{1}{\sqrt{M}}\frac{|\sin(\pi\psi_l M)|}{|\sin(\pi\psi_l )|}$
is a localized kernel around $\theta_l=\sin^{-1}[\frac{2(m'-1)}{M}-1]$ with a resolution of $\frac{1}{M}$.
In general, the AoA-AoDs of the MPCs are not aligned with the discrete grid $\clG=\Theta \times \Phi$. However, as the number of antennas $M$ at the BS and $N$ at the UE increases, the DFT basis provide \textit{good sparsification} of the channel matrix $\check{\bfH}_{s}[\omega]$. 
This is qualitatively illustrated in Fig.\,\ref{sparsity} for a channel with $L=3$ discrete MPCs and off-grid AoA-AoD components. 
It is seen that, as $M$ and $N$ increase, the resulting representation $\check{\bfH}_{s}[\omega]$ is more and more sparse.


\section{Proposed Beam-Alignment Algorithm}\label{Mathematical}

\subsection{High-Level Overview}\label{High-Level Overview}
\begin{figure}[t]
	\centerline{\includegraphics[width=8cm]{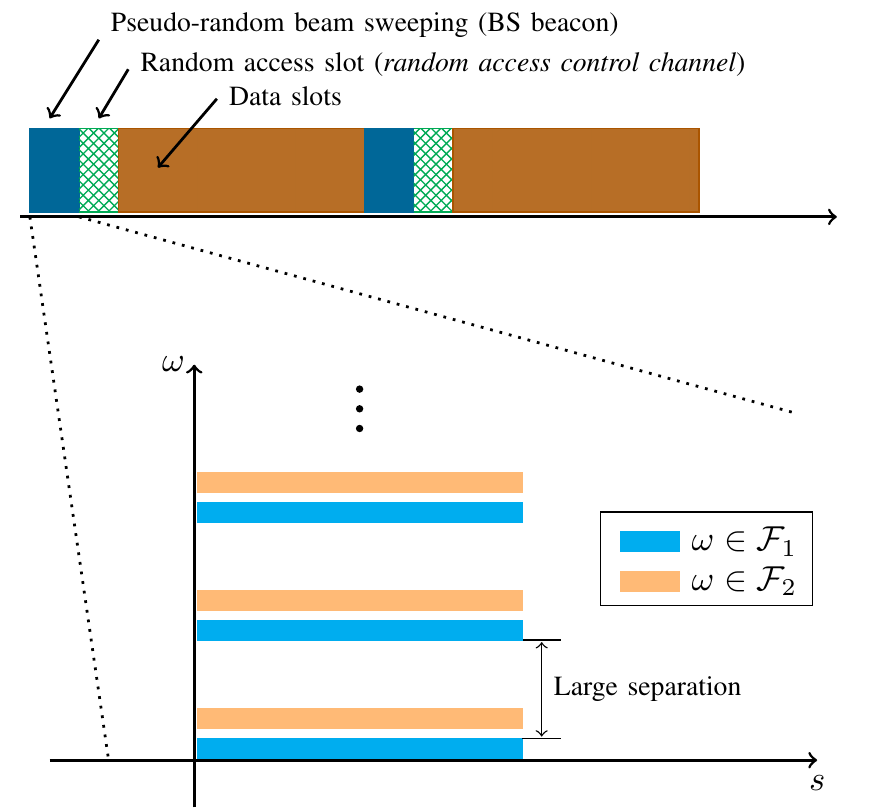}}
	\caption{{\small {\em (Top) Frame structure of the proposed BA scheme. 
				(Bottom) Different beacon signals are orthogonally multiplexed over disjoint sets of subcarriers $\omega\in\clF_i$, $i\in[m]$. 
				In the figure's example we have two orthogonal beacon signals on the ``blue'' and on the ``orange'' combs of subcarriers}}.}
	\label{Fig:frameStructure}
\end{figure}
In this section, we provide a high-level overview of our scheme, the details of which are 
presented in the next sections. 
Fig.~\ref{Fig:frameStructure} shows the proposed frame structure.
During the {\em DL beacon slots}, the BS broadcasts a probing signal to all the UEs, while the UEs stay in listening mode. The probing signal is formed by a sequence of pseudo-random beam patterns (referred to as the transmit beamforming codebook),  repeated periodically, and priori known to all users. Each UE makes measurements of the beacon transmission by applying its own (individual) sequence of receive beam patterns (referred to as the receive beamforming codebook) during the beacon slots. The number of measurements may differ from user to user, depending on the individual pre-beamforming SNR and on the number of receiver RF chains. 
When a UE has gathered enough measurements, it estimates the AoA-AoD information for its strongest MPC (using the method detailed in the following), 
and transmits over the UL {\em random access slot} a beamformed control packet using a beamforming vector in the direction of its best estimated AoA (UE side). The control packet contains the user ID, the index of the estimated best AoD (BS side), 
and possibly some additional protocol information (e.g., rate information which can be derived from the estimated SNR after beamforming). 
During the random access slots, the BS stays in listening mode and uses its $m$ RF chains to form $m$ coarse beam patterns (sectors), 
covering the whole BS angle domain, in order to provide some receiver beamforming gain. 
We assume that, when the UE has correctly estimated its best MPC, with high probability, the beamforming gain at the UE side and the sectorization at the BS side are sufficient to successfully decode the control packet.  
This assumption is justified by the fact that the control packet can be encoded at low rate. 
After sending the UL control packet, the UE puts itself in listening mode with the same beamforming vector used to send the control packet in the UL.  At this point, the BS  knows the beamforming vectors to be used for all UEs whose control packet was successfully decoded. Each time, after BS decodes the control packet of a new connecting user, the BS responds with a beamformed \textit{Acknowledgment} (ACK) packet sent in one of the data slots.  It follows that a beamforming gain at both the BS and the UE side is achieved for the ACK packet 
and for all subsequent data packets, both in the DL and in the UL.  Therefore, data communication can take place at a rate that depends 
on the SNR after beamforming.\footnote{Once BA is achieved and regular data communication
takes place, the BS and the UE can keep using the beacon slot in addition to the data slots to enable some
beam tracking algorithm and keep the alignment over smooth variations of the AoA-AoDs. In this paper we are only concerned with the {\em initial} BA, 
i.e., when the UE needs to connect to the BS without {\em a priori} information on the direction of its strong MPCs.}
In the case that the control packet is received in error (e.g., this may be due to an estimation error in the AoA/AoD information, or to a collision in the random access slot), the BS can not respond with a correct beamformed ACK, and the UE, after a given time-out, will realize that something went wrong. Then, the UE keeps gathering more measurements from the beacon slots, and the BA procedure restarts.

\subsection{BS Channel Probing and UE Sensing}\label{probing and sensing}

Without loss of generality, we focus on the BA procedure for a generic UE and omit the UE index, 
since the scheme applies in parallel and virtually without mutual interaction for all connecting UEs (apart from possible collisions in the random access UL slot). 
Consider the channel matrix $\bfH_{s}[\omega]$ between the BS and the UE arrays, as defined  in Section \ref{Quantization}, 
and its beamspace representation  $\check{\bfH}_{s}[\omega]$ at beacon slot $s\in [T]$ and subcarrier $\omega \in[F]$,
where $T$ is the effective period of beam training. 

For simplicity of exposition, we assume that the beacon slot contains a single OFDM symbol interval.\footnote{The generalization to 
multiple OFDM symbols per slot is immediate and slots of $S\geq 1$ OFDM symbols shall be used in the numerical results.}
At each beacon slot, the BS uses its $m$ RF chains to probe the channel along $m$ beamforming vectors $\bfu_{s,i}$, $i \in [m]$,  
by transmitting an OFDM symbol $x_{s,i}(t)$ along each $\bfu_{s,i}$. We design the beacon OFDM symbols 
$x_{s,i}(t)$ such that they are mutually orthogonal in the frequency domain. 
In particular, for each $i \in [m]$ we define a subset $\clF_i \subset[F]$ of size $|\clF_i| \leq F$ 
such that $\clF_i \cap \clF_{i'}=\emptyset$ for $i\neq i'$
(see Fig.\,\ref{Fig:frameStructure}).  We make the additional assumption that each subset $\Fc_i$ forms a ``comb'' of equal size $|\Fc_i| = F'$, 
with widely separated subcarriers such that the frequency separation is larger than the channel coherence bandwidth. Hence, 
the corresponding channel matrices $\bfH_{s}[\omega]$ for the different subcarriers $\omega \in  \clF_i$ are mutually uncorrelated.

A main ingredient of our proposed BA scheme is the pseudo-random beamforming codebook 
transmitted by the BS during the beacon slots, defined as the collection of sets $\clC_\text{BS}:= \{\Uc_{s,i} : s \in [T], i \in [m] \}$, 
where $\Uc_{s,i}$ is the angle-domain support (i.e., the subset of quantized angles in the virtual beamspace representation) 
defining the directions to which the transmit beam patterns $\uv_{s,i}$ sends the signal power. 
We let $|\Uc_{s,i}| = \kappa_u \leq M$ for all $(s,i)$. The beamforming vectors are given by $\uv_{s,i} = \Fm_M \check{\uv}_{s,i}$, where
$\check{\bfu}_{s,i}=\frac{\one_{\clU_{s,i}}}{\sqrt{\kappa_u}}$, and where $\one_{\clU_{s,i}}$ denotes a vector with $1$ 
at components in the support set  $\clU_{s,i}$ and $0$ elsewhere.  
An example of such patterns with the corresponding vector $\check{\bfu}_{s,i}$ is shown in Fig.\,\ref{cluster} (a). 
The pseudo-random nature of the codebook is  due to the fact that the sequences of angular support sets $\{\Uc_{s,i} : i\in [m], s \in [T]\}$ 
are generated in a pseudo-random manner. 

The second ingredient of our proposed BA algorithm is a local receive codebook 
at each UE, through which the UE makes measurements in order to 
estimate the AoA-AoD information of its strong MPCs.  Each UE can customize (locally) its own receive beamforming codebook  defined by 
the collection of sets $\clC_\text{UE}:= \{ \Vc_{s,j} : s \in [T], j \in [n]\}$, where  
 $\Vc_{s,j}$ is the angle-domain support defining the directions from which the receiver beam 
 patterns $\vv_{s,i}$ collect signal power.  We let $|\Vc_{s,j}| = \kappa_v \leq N$ for all $(s,j)$. The beamforming vectors are given 
 by $\vv_{s,j} = \Fm_N \check{\vv}_{s,j}$, where $\check{\bfv}_{s,j}=\frac{\one_{\clV_{s,j}}}{\sqrt{\kappa_v}}$. 
 Similar to the parameter $\kappa_u$ at the transmitter side, the parameter $\kappa_v$ controls the spread of the sensing window at the UE side. 
This is illustrated again in Fig.\,\ref{cluster} (a).

During the $s$-th beacon slot, the UE applies the receive beamforming vector $\bfv_{s,j}$ to its $j$-th RF chain, 
obtaining the frequency-domain received signal (after OFDM demodulation) given by (\ref{receiveblock}) for 
$i \in [m]$ and $\omega \in \Fc_i$.  Note that the $m$ probing signals $x_{s,i}(t)$ are orthogonal in the frequency domain and therefore can be perfectly separated at the receiver.  It is convenient to write \eqref{receiveblock} directly in terms of the beamspace representation as 
\begin{equation}\label{receiveblock_quant}
\check{y}_{s,i,j}[\omega]=\frac{1}{\sqrt{n}} \check{\bfv}_{s,j}^\herm \check{\bfH}_{s}[\omega]\check{\bfu}_{s,i} \check{x}_{s,i}[\omega]+\check{z}_{s,j}[\omega], \;\;\;\; \omega \in  \clF_i.
\end{equation}
The BS total transmit power $\ptot$ is allocated equally to  all the probing streams $i \in [m]$, all the subcarriers in $\omega \in \Fc_i$, and all the
 $\kappa_u$ beamspace directions. Hence, the symbols $\{\check{x}_{s,i}[\omega] : \omega \in \clF_i\}$ 
 have uniform power distribution with $\bE[|\check{x}_{s,i}[\omega]|^2] = \frac{\ptot}{m F'} := \pdim$ (power per transmit signal dimension).  
 In fact, without loss of generality,  we choose the frequency-domain probing symbols to be constant and given by $\check{x}_{s,i}[\omega]= \sqrt{\pdim}$. 


For the class of beamforming patterns defined by $\clC_\text{BS}$ and by $\clC_\text{UE}$, 
from (\ref{loc_kernel}) it is easy to show that $|\gBS_{s,l,i}|^2 \leq M/\kappa_u$ and $|\gUE_{s,l,j}|^2 \leq N/\kappa_v$. 
It follows that
\begin{equation} 
\bE \left [ \left |  \check{\bfv}_{s,j}^\herm \check{\bfH}_{s}[\omega]\check{\bfu}_{s,i}\right  |^2\right ] \leq \frac{M N \sum_{\ell =1}^L \gamma_\ell}{\kappa_u \kappa_v}.
\end{equation} 
Using this bound in (\ref{snraft_co-equal}), we obtain the maximum possible SNR for channel estimation in the per-subcarrier observation (\ref{receiveblock_quant}), given by 
\begin{align}
\snraftce &:= \frac{\pdim}{n} \frac{MN \sum_{l=1}^L \gamma_l}{\kappa_u \kappa_v \sigma^2} \nonumber\\
&= \frac{MN}{\kappa_u \kappa_v mn}  \times \frac{B}{F' \Delta f} \times \snrbefco. \label{snrafbce}
\end{align}

Expression (\ref{snrafbce}) puts in evidence the role of the different factors: the first term is the ratio of the maximal available beamforming gain
$MN$, divided by the total signal dimensions in the spatial multiplexing domain $\kappa_u \kappa_v mn$. 
The second term corresponds to the power concentration in the frequency domain, and the third term 
is the SNR before beamforming, defined in (\ref{snrbef_co}).


The frequency spreading factor  $F'$ and angle spreading factors $\kappa_u,\kappa_v$ can be optimized depending on the specific cell topology
(e.g., on the size of the cell, which in turns determines the worst-case SNR before beamforming).  
Clearly, by making $\kappa_u$ (resp., $\kappa_v$) larger, each beam pattern probes (resp., sense) 
simultaneously more directions, but the total power is spread over all such directions. 
In contrast, by making $\kappa_u$ (resp., $\kappa_v$)  smaller, the beam pattern explores less directions
but obtains better power concentration in the angle domain. It is also important to notice the effect of $F'$: as we shall see in Section \ref{estimator}, 
the AoA-AoD estimator builds some sample-mean statistics by averaging over a sufficiently large number of uncorrelated
channel fading realization over the frequency domain. Hence, larger $F'$ provide better averaging at the cost of spreading the total power 
over more subcarriers.\footnote{This tradeoff in 
the choice of the spreading parameters $F'$ and $\kappa_u, \kappa_v$ can be seen as an instance of the well-known \textit{exploration-exploitation} tradeoff in statistics.} 

\begin{remark}
	The proposed BS pseudo-random beamforming codebook can be regarded as a generalization of the classical ``coarse beam sweeping'', 
	where the probing vectors pack all directions in the given coarse angle intervals (sectors) \cite{alkhateeb2014channel, KokOverlap}. 
	Our results show that using the proposed method with
	random beam patterns instead of the traditional coarse beam sweeping, we can achieve a very good beam alignment without the need of the
	interactive {\em beam refinement} phase as usually considered in current practical approaches (e.g., see \cite{Nitsche2014Overview}). 
\end{remark}

\begin{remark}
In the proposed scheme, the channel probing takes place only in the DL, where BS actively probes the channel while all the UEs are in the receiving mode. 
Thus, our scheme is markedly different from interactive channel estimation schemes in which both the UE and the BS take turns and probe the channel in a bi-directional way (e.g., see \cite{xiarobust2,Nitsche2014Overview,tsang2011coding}). 
In general, interactive schemes require some coordination among the UEs, which is difficult to establish in the initial acquisition mode. 
In the absence of such a coordination, the simultaneous transmission of those UEs interested in joining the system might create severe 
multi-access interference to the UEs already in the system.  
\end{remark}

\subsection{Channel Estimation at the UE Side}   \label{estimator}

The strong MPCs of the channel correspond to the 
components $(k,k')$ in the matrix $\check{\Hm}_s[\omega]$ with large second moment. Notice also that the element second moments $\bE[ |(\check{\Hm}_s[\omega])_{k,k'} |^2]]$ are invariant both with respect to $s$ (time) and with respect to $\omega$ (frequency). This follows immediately from the channel model definition and the assumption of uncorrelated MPCs. If we had direct access to measurements of the elements of $\check{\Hm}_s[\omega]$,  a naive approach would build estimators for the second moments (as sample mean), and try to identify the largest. However, this would require a number of RF chains equal to the number of antenna elements. In contrast, we have only access to the projections $\check{\bfv}_{s,j}^\herm \check{\bfH}_{s}[\omega]\check{\bfu}_{s,i}$ from the observation in (\ref{receiveblock_quant}). 

Using  $\check{x}_{s,i}[\omega] = \sqrt{\pdim}$ in (\ref{receiveblock_quant}), we can write the received beacon symbol observation at the UE  as
\begin{align}\label{receiveblock_quant2}
\check{y}_{s,i,j}[\omega]
&=\sqrt{\frac{\pdim}{n}} \check{\bfv}_{s,j}^\herm \check{\bfH}_{s}[\omega]\check{\bfu}_{s,i} +\check{z}_{s,j}[\omega] \nonumber \\
&=\sqrt{\frac{\pdim}{n}} (\check{\bfu}^\transp_{s,i}\otimes \check{\bfv}^\herm_{s,j}) \check{\bbh}_{s}[\omega] +\check{z}_{s,j}[\omega] \nonumber \\
&=\sqrt{\frac{\pdim}{n}}\, \bfg_{s,i,j}^\herm \check{\bbh}_{s}[\omega] +\check{z}_{s,j}[\omega],
\end{align}
where $\check{\bbh}_s[\omega]=\vec(\check{\bfH}_{s}[\omega])$ denotes the vectorized beamspace representation of the 
channel matrix  at subcarrier $\omega \in \clF_i$, where we used the well-known identity $\vec(\bfA \bfB \bfC)=(\bfC^\transp \otimes \bfA) \vec(\bfB)$, 
and where we define the combined probing and sensing beamforming pattern as  $\bfg_{s,i,j}=\check{\bfu}^*_{s,i}\otimes \check{\bfv}_{s,j} \in \bC^{MN}$, which is common across all the subcarriers $\omega \in \clF_i$ but differs for different pairs of RF chains $(i,j)$ at the BS and the UE.

In practice, each beacon slot is formed by a block of $S \geq 1$ OFDM symbols. With a slight abuse of notation, 
we index the symbols belonging to the $(s+1)$-th slot as $sS + s'$, for $s' \in [S]$. 
The instantaneous received power at the $j$-th RF chain of the UE from the signal transmitted by the $i$-th RF chain of the  BS on the $s$-th beacon slot
is given by 
\begin{align}\label{Q(t)}
\check{q}_{s,i,j} &= \frac{1}{SF'}\sum_{s' \in [S]} \sum_{\omega \in \clF_i}|\check{y}_{sS+s',i,j}[\omega]|^2\nonumber \\
&=\frac{\pdim}{n} \bfg_{s,i,j}^\herm   \left ( \frac{1}{S F'} \sum_{s' \in [S]} \sum_{\omega \in \clF_i}\check{\bbh}_{sS+s'}[\omega]\check{\bbh}_{sS+s'}[\omega]^\herm \right ) \bfg_{s,i,j} \nonumber \\
& +\frac{1}{SF'} \sum_{s' \in [S]} \sum_{\omega \in \clF_i} |\check{z}_{sS+s',j}[\omega]|^2 + \frac{1}{SF'} \sum_{s' \in [S]}  \sum_{\omega \in  \clF_i} {\xi}_{sS+s',j}[\omega],
\end{align}
where the first and the second term correspond to the signal contribution and to the noise contribution, and where 
\begin{align*}
{\xi}_{sS+s',j}[\omega]=2 \sqrt{\frac{\pdim}{n}} \Re\left \{ \bfg_{s,i,j}^\herm \check{\bbh}_{sS+s'}[\omega] \check{z}_{sS+s',j}[\omega]^\herm \right \} 
\end{align*}
denotes the signal-noise cross term. 
The key idea underlying our method follows from the fact that, when the number of dimensions $S \times F'$ over which the averaging of the instantaneous power
is large, such that the signal-noise  term becomes negligible and the empirical covariance matrix of the channel vector converges as
\begin{align}
\frac{1}{SF'} \sum_{s' \in [S]} \sum_{\omega \in \clF_i}\check{\bbh}_{sS+s'}[\omega]\check{\bbh}_{sS+s'}[\omega]^\herm \to \bE[ \bbh_s[\omega] \bbh_s[\omega]^\herm] =:\Sigmam_{\bbh}.
\end{align}
Meanwhile the noise term converges to
\begin{align}
\frac{1}{SF'} \sum_{s' \in [S]} \sum_{\omega \in \clF_i}|\check{z}_{sS+s',j}[\omega]|^2 \to \sigma^2.
\end{align}
The received power in \eqref{Q(t)} gives a 1-dimensional noisy projection of the 
covariance matrix $\Sigmam_{\bbh}$ with respect to the combined probing and sensing vector $\gv_{s,i,j}$.  
It is important to note that $\Sigmam_{\bbh}$ is independent of the subcarrier and time slot indices $\omega$ and $s$, respectively, due to the channel stationarity in the frequency and time domain.

In order to derive our proposed estimation method, we approximate \eqref{Q(t)} as 
\begin{align}\label{dumm1}
\check{q}_{s,i,j}\approx \frac{\pdim}{n} \bfg_{s,i,j}^\herm \Sigmam_{\bbh} \bfg_{s,i,j} + \sigma^2.
\end{align}
When all the AoA-AoDs lie on a discrete grid, $\check{\bbh}_s[\omega]$ is a sparse vector with i.i.d. components with only a few nonzero coefficients corresponding to the scatterers. Due to the independence of the channel gain of the scatterers, the covariance matrix of $\check{\bbh}_s[\omega]$ would be a diagonal $MN\times MN$ matrix with only a few nonzero diagonal elements corresponding to the scatterers. 
In practice, $\Sigmam_{\bbh}$ is still sparse and approximately diagonal for sufficiently large $M$ and $N$ (as illustrated in Fig.\,\ref{sparsity}), 
even if  the AoA-AoDs of the scatterers do not lie on the discrete grid.

\begin{figure*}[t]
	\centering
	\includegraphics{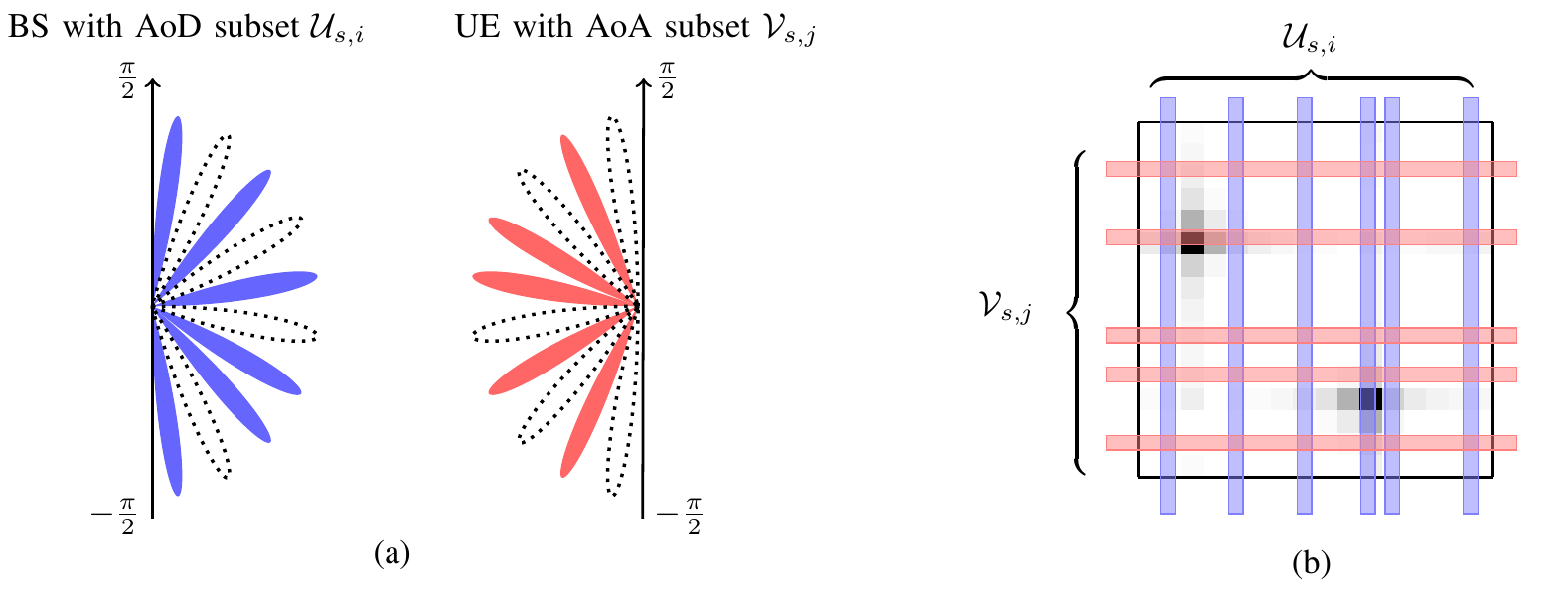}
	\caption{{\small {\em (a) Illustration of the subset of AoA-AoDs at time slot $s$ probed by the $i$-th beacon stream
	transmitted by the BS and received by the $j$-th RF chain of the UE, for $M = N = 10$. 
	The AoD subset is given by  $\clU_{s,i}=\{1,3,4,6,8,10\}$ with beamforming vector 
	$\check{\bfu}_{s,i}=\frac{1}{\sqrt{6}}[1,0,1,0,1,0,1,1,0,1]^\transp$. The AoA subset is given by 
$\clV_{s,j}=\{2,4,5,7,9\}$ with receive beamforming vector $\check{\bfv}_{s,j}=\frac{1}
{\sqrt{5}}[0,1,0,1,1,0,1,0,1,0]^\transp$.
(b) the channel gain matrix $\check{\Gammam}$ (with two strong MPCs indicated by the dark spots) 
measure along $\clV_{s,j} \times \clU_{s,i}$}}.}
	\label{cluster}
\end{figure*}

Using the expression for the probing and sensing vectors $\check{\bfu}_{s,i}=\frac{\one_{\clU_{s,i}}}{\sqrt{\kappa_u}}$ 
and $\check{\bfv}_{s,j}=\frac{\one_{\clV_{s,j}}}{\sqrt{\kappa_v}}$, respectively, \eqref{dumm1} can be further simplified to
\begin{align}\label{dumm2}
\check{q}_{s,i,j}\approx \frac{\pdim M N}{n \kappa_u \kappa_v} \sum_{(r,c) \in \clN_{s,i,j}} \check{\Gamma}_{r,c}   + \sigma^2,
\end{align}
where $\clN_{s,i,j}=\clV_{s,j}\times \clU_{s,i}$ denotes the subset of AoA-AoD quantized directions 
probed/sensed by the beamforming vectors $\check{\bfu}_{s,i}$ and $\check{\bfv}_{s,j}$, respectively, 
and where $\check{\Gamma}_{r,c} = \bE[|[\check{\Hm}_s[\omega]]_{r,c}|^2]$ denotes the 
second moment of the channel coefficient of the scatterer located at the discrete AoD $c$ at the BS and AoA $r$ at the UE. 
Notice that because of the sparsity of the channel in the virtual beam domain, most elements $\check{\Gamma}_{r,c}$ are equal to or near zero, 
and only a few, corresponding to the directions $(r,c)$ strongly coupled by scatterers, are large. 
With reference to Fig.\,\ref{cluster} (b), letting $\check{\Gammam}$ denote the $N\times M$ matrix with elements $\check{\Gamma}_{r,c}$, 
we notice that the summation in (\ref{dumm2}) includes all elements $\check{\Gamma}_{r,c}$ at the index coordinates $(r,c) \in \clN_{s,i,j}$ 
at the crossing points of the probing directions (vertical lines in Fig.\,\ref{cluster} (b)) and the sensing directions (horizontal lines in 
Fig.\,\ref{cluster} (b)). Hence, 
the term $\frac{\pdim M N}{n \kappa_u \kappa_v} \sum_{(r,c) \in \clN_{s,i,j}} \check{\Gamma}_{r,c} $ in 
(\ref{dumm2}) can be further written as the inner product $\bfb_{s,i,j}^\transp \gammam$ where
$\bfb_{s,i,j} := \one_{\clU_{s,i}}\otimes \one_{\clV_{s,j}}$ is a binary vector containing $1$ at the AoA-AoDs probed 
by $\clV_{s,j} \times \clU_{s,i}$ and is $0$ elsewhere, and where we define
$\gammam :=   \frac{\pdim M N}{n \kappa_u \kappa_v}   \vec(\check{\Gammam}) \in \bR_+^{MN}$. 
In general, for a finite number of subcarriers, \eqref{dumm1} -- (\ref{dumm2}) hold only approximately  
since the statistical fluctuations are not negligible.  We consider this as a residual noise $\check{w}_{s,i,j}$, such that 
the approximation in \eqref{dumm2} yields the equality
\begin{align}\label{dumm4}
\check{q}_{s,i,j} = \bfb_{s,i,j}^\transp \gammam + \sigma^2 + \check{w}_{s,i,j}.
\end{align}
In each beacon slot, since the BS transmits along $m$ RF chain in each beacon slot and the UE has $n$ RF chains to sense the channel, 
the UE obtains $mn$ equations for the unknown vector $\gammam$ as in \eqref{dumm4}. 
Over $T$ beacon slots the UE obtains $mnT$  equations, which can be written in the form
\begin{align}\label{UE_equations}
\check{\bfq}=\bfB \gammam + \sigma^2 \one + \check{\bfw},
\end{align}
where the vector ${\check{\bfq}=[\check{q}_{1,1,1}, \dots \check{q}_{1,m,n}, \dots, \check{q}_{T,1,1}, \dots, \check{q}_{T,m,n}]^\transp}$ consists of all $mnT$ measurements calculated as in (\ref{Q(t)}),  where the $mnT \times MN$  matrix ${\bfB=[\bfb_{1,1,1}, \dots, \bfb_{1,m,n}, \dots, \bfb_{T,1,1}, \dots, \bfb_{T,m,n}]^\transp}$ is uniquely defined by the beamforming codebooks $\Cc_{\rm BS}$ and $\Cc_{\rm UE}$, 
and where $\check{\bfw} \in \bR^{mnT}$ is the residual noise in the measurements. 
At this point, some remarks are in order.

\begin{remark}
In our proposed scheme, at each acquisition slot, each UE extracts its own set of measurements from its received signal.  An implicit assumption, however, is that each UE is synchronized with the BS and knows the BS codebook $\clC_\text{BS}$ such that it can construct the 
matrix $\Bm$ in \eqref{UE_equations}. This assumption is explicitly or implicitly made in virtually all
works dealing with initial beam acquisition (aka, BA problem), as reviewed in Section \ref{introduction}. Therefore, this is not a particularly restrictive assumption
specific to our approach.  
\end{remark}

\begin{remark}
While the BS codebook $\clC_\text{BS}$ is  broadcasted to all UEs,  the sensing codebook $\clC_\text{UE}$
can be user-specific. In particular, we may imagine a system where each UE has a set of possible codebooks, 
characterized by a different number of (active) receive RF chains $n$ and sensing directions $\kappa_v$, and uses the most appropriate codebook depending on its hardware and value of SNR before beamforming. Intuitively, \textit{strong} users (suffering from a small pathloss) can select larger $\kappa_v$ and/or $n$ to better explore the channel directions and estimate $\gammam$ faster, 
whereas \textit{weak} users (suffering form a large pathloss) should select a smaller $\kappa_v$ and/or $n$ to attain a reasonable training SNR (see \eqref{snrafbce}). Although the  \textit{weak} users might need to wait longer to take more measurements before they are able to estimate their channels, \eqref{UE_equations} remains still valid since the channel gains (second order channel statistics) $\gammam$ are stable across many channel coherence blocks. This is in stark contrast with the conventional CS-based techniques used for BA via estimating the instantaneous complex channel gains, where the underlying channel might change drastically while taking the measurements, especially when only very few number of RF chains $m,n$ are available. Thus,  in these schemes, $T$ is limited by the channel coherence time and this prevents from accumulating enough measurements and enough signal power. This effect is clearly visible in the results of Section \ref{performance}, where we compare our method with state-of-the-art CS-based methods. 
\end{remark}

\subsection{Non-Negative Least Squares}

In order to identify the AoA-AoD directions of the strong scatterers, we estimate the $MN$ dimensional vector 
$\gammav$ from the $mnT$-dimensional observation  given in \eqref{UE_equations}. 
Because of the presence of the measurement noise $\check{\bfw}$, a standard approach consists of solving the 
Least-Squares (LS) problem $\min_{\gammam} \|\bfB \gammam + \sigma^2 \one - \check{\bfq}\|^2$. However,  in general 
$MN$ is significantly larger than $mnT$, such that the system of equations is heavily underdetermined and 
the LS solution yields meaningless results. 
The key observation here is that $\gammav$ is {\em sparse} (by assumption) and {\em non-negative} (by construction).  
%
Recent results in CS show that when the underlying parameter $\gammam$ is non-negative, 
the simple non-negative constrained LS given by
\begin{align}\label{eq:NNLS}
\gammam^*=\argmin_{\gammam \in \bR_+^{MN}} \|\bfB \gammam + \sigma^2 \one - \check{\bfq}\|^2, 
\end{align}
is still enough to impose sparsity of the solution $\gammam^*$ \cite{slawski2013non, RN276}, with no need for an explicit 
sparsity-promoting regularization term in the objective function as in the classical LASSO algorithm \cite{tibshirani1996regression}.
The (convex) optimization problem \eqref{eq:NNLS} is generally referred to as \textit{Non-Negative Least Squares} (NNLS), 
and has been well investigated  in the literature. 
An early reference is \cite{donoho1992maximum}, showing that NNLS might yield a ``Super-Resolution'' property
depending on the structure of the measurement matrix (e.g., $\bfB$ in our case).   
More recently, by the advent of CS \cite{donoho2006compressed,candes2006near}, the NNLS has reemerged in the context 
of sparse signal recovery, where it has been shown that the non-negativity constraint alone might suffice to recover the underlying 
signal $\gammam$ in the noiseless  \cite{bruckstein2008uniqueness, donoho2010counting, wang2009conditions, wang2011unique} as well as 
in the noisy case \cite{slawski2013non, RN276}. Moreover, \cite{slawski2013non} demonstrates that NNLS has a noisy recovery performance 
comparable to that of LASSO. In \cite{slawski2013non} it is also shown that  NNLS along with an appropriate thresholding 
provides state-of-the-art performance in terms of support estimation. This property is very relevant in the context of 
this paper, where the identification of the support of $\gammav$ corresponds to finding the AoA-AoD directions 
strongly coupled by MPC. 

In terms of numerical implementations, the NNLS can be posed as an unconstrained  LS problem over the positive orthant 
and can be solved  by several efficient techniques such as Gradient Projection, Primal-Dual techniques, etc., with an affordable computational complexity \cite{bertsekas2015convex}. We refer to \cite{kim2010tackling, nguyen2015anti} for the recent progress on the numerical solution of 
NNLS and  a discussion on other  related work in the literature. 

\section{Performance Evaluation}\label{performance}

In this section we evaluate the the performance of our proposed algorithm via numerical simulations. 
To run the NNLS optimization in \eqref{eq:NNLS}, we use the  implementation of NNLS in\ \matlab called \texttt{lsqnonneg.m}.

{\bf Channel and Signal Model.}
For simplicity, we consider a very sparse channel model with only one path (one scatterer). 
We assume $f_0 = 70$ GHz carrier frequency and bandwidth of $B=1$ GHz. The OFDM subcarrier spacing is $480$ kHz 
in compliance with recent 3GPP standard specifications \cite{3GPP2017}. Assuming $\tau_{\rm cp} \Delta f = 0.25$ (i.e., the CP length is 25\% of the OFDM duration)   we obtain  $t_0 = 2.6$ $\mu$s and around $F = 2048$ subcarriers (plus some guard band). 
We fix the frame duration of our scheme (i.e., the  repetition interval of the beacon slot) to $1$ ms, consists of  $384$ OFDM symbols (per subcarrier). 
A \textit{beacon slot} contains $S = 14$ OFDM symbols, the \textit{random access slot} also contains $14$ OFDM symbols, and the 
remaining $356$ symbols are used for data transmission \cite{3GPP2017}. Unless otherwise specified, we assume that the BS has $M=32$ antennas and  $m=3$ RF chains, 
and the UE has $N=32$ antennas and $n=2$ RF chains. The simulations in this paper consider $\snrbef=-33$ dB (unless otherwise stated) \cite{3GPP2017}. 
We announce an individual experiment to be successful if the index of the strongest component in $\gammam$ 
is correctly estimated (i.e., it coincides with the actual scatterer location, up to the discrete angle grid quantization).

{\bf Dependence on the Random BS Codebook.}
We generated at random $4$ different probing codebooks at the BS side. 
Fig.\,\ref{changecodebook} illustrates the detection probability for the different pseudo-random codebooks, 
where the power spreading factors at the BS and the user sides are set to {$\kappa_u = \kappa_v = 16$} respectively. 
We repeated each experiment $200$ times and plot the resulting detection probability versus training period length $T$. 
As expected, increasing $T$ significantly improves the detection probability. More importantly, different codebooks have quite similar performances. 
This shows the fact that the performance of our scheme is quite insensitive to the choice of the random probing codebook, as long as 
it is sufficiently randomized.

\begin{figure}[t]
	\centering
	\includegraphics[width=8cm]{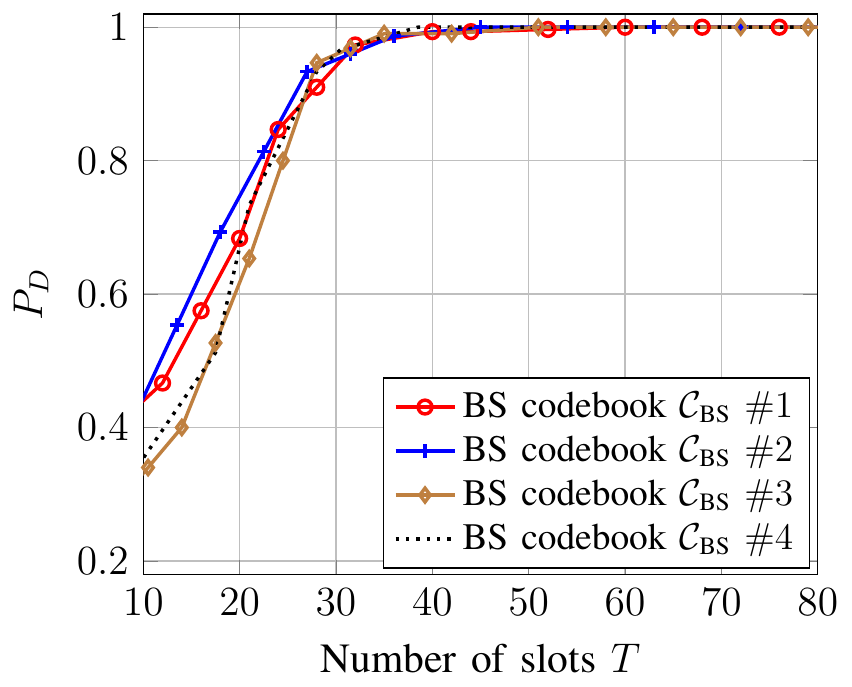}
	\caption{{\small {\em Comparison of the detection probability $P_D$ of different pseudo-random codebooks (denoted by $\clC_\text{BS}$) achieved by the proposed NNLS scheme, for $M =N=32$, $F'=3$, $m=3$, $n=2$, $\snrbef=-33$ dB, $\kappa_u = \kappa_v = 8$}}.}
	\label{changecodebook}
\end{figure}

{\bf Dependence on the Beam Spreading Factors $\kappa_u$ and $\kappa_v$.}
The spatial spreading factors $\kappa_u$ and $\kappa_v$ impose a trade-off between the  angle coverage  of the probing/sensing 
matrix $\bfB$ (exploration) and its receive SNR at the user side (exploitation).  This is illustrated in Fig.\,\ref{changeKVKU}. 
It is seen that increasing the spreading factor from $\kappa_u=\kappa_v=4$ to $\kappa_u=\kappa_v=8$ improves the performance. 
However, increasing $\kappa_u$, $\kappa_v$ to $\kappa_u=\kappa_v=16$ slightly degrades the performance, 
and the degradation is severe when $\kappa_u$, $\kappa_v$ are increased to $\kappa_u=\kappa_v=25$.

\begin{figure}[t]
	\centering
	\includegraphics[width=8cm]{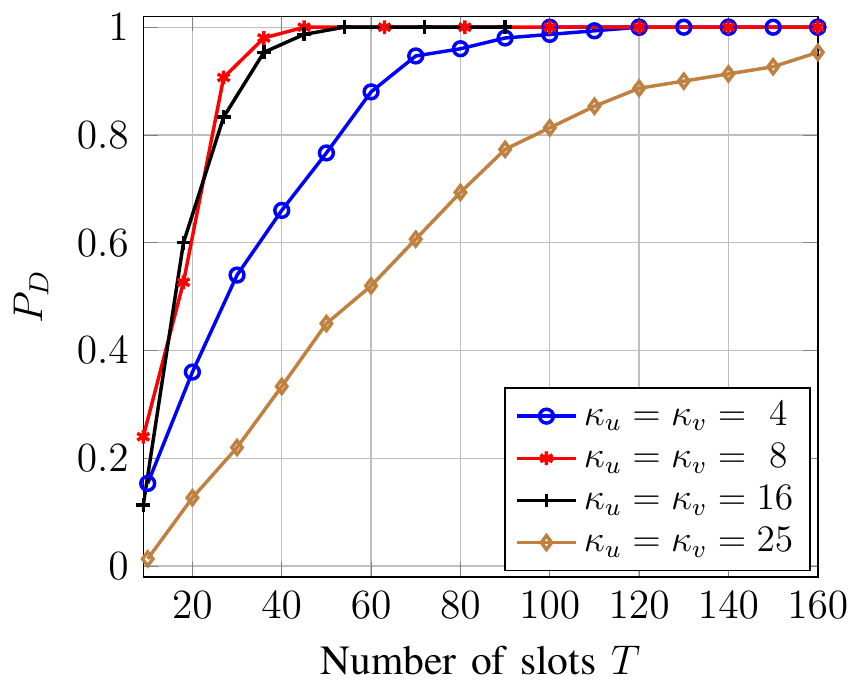}
	\caption{{\small {\em Detection probability $P_D$ of our proposed scheme for different power spreading factors ($\kappa_u$, $\kappa_v$), where $M=N=32$, $F'=3$, $m=3$, $n=2$, $\snrbef=-33$ dB}}.}
	\label{changeKVKU}
\end{figure}

{\bf Dependence on the number of subcarriers $F'$.}
As explained in Section \ref{probing and sensing} and \ref{estimator}, a large number of subcarriers $F'$ deployed for each RF chain at the BS side (together with the factor $S$, i.e., the number of OFDM symbols contained in a slot) ensures a reliable averaging of the instantaneous power received at the UE side. This averaging scheme makes the residual noise term in \eqref{dumm4} approximately negligible, hence, ensuring a good performance of the NNLS estimator. However, increasing $F'$ means a larger spreading of the total power. As shown in Fig.\,\ref{changeF}, increasing the number of subcarriers from $F'=1$ to $F'=3$ improves the performance, but increasing  $F'$ to $F'=10, 30$ degrades the performance considerably.

\begin{figure}[t]
	\centering
	\includegraphics[width=8cm]{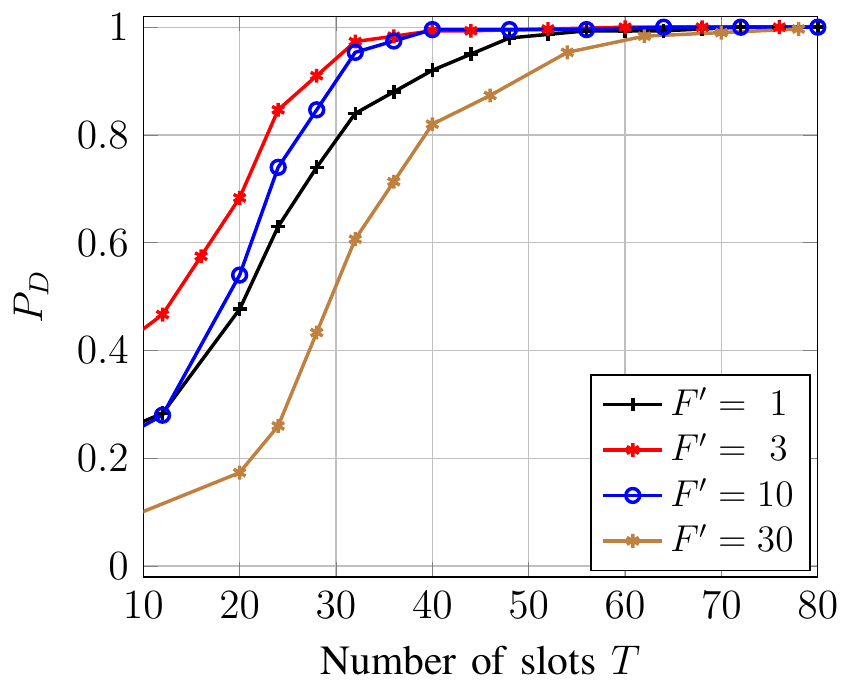}
	\caption{{\small {\em Detection probability $P_D$ of our proposed scheme for different number of subcarriers $F'$ deployed for each RF chain at the BS side, where $M=N=32$, $m=3$, $n=2$, $\snrbef=-33$ dB}}.}
	\label{changeF}
\end{figure}

{\bf Dependence on the probing dimensions $(\kappa_u\kappa_v mn)$.}
Note that, for a certain pre-beamforming SNR, the output of the proposed BA scheme inherently depends on the probing dimensions, i.e., the product $\kappa_u\kappa_v mn$. This is illustrated in Fig.\,\ref{Constant_mnkvku}. When $(\kappa_u\kappa_v mn)$ are constant, the performance of the proposed scheme is independent of the specific values of $\kappa_u$, $\kappa_v$, $m$, $n$. Hence, as far as BA is concerned, 
one can reduce the UE hardware complexity (number of RF chains) by either adding more RF chains at the BS, or increasing the spatial 
spreading factor $\kappa_u$ and/or $\kappa_v$, while achieving a similar performance.

\begin{figure}[t]
	\centering
	\includegraphics[width=8cm]{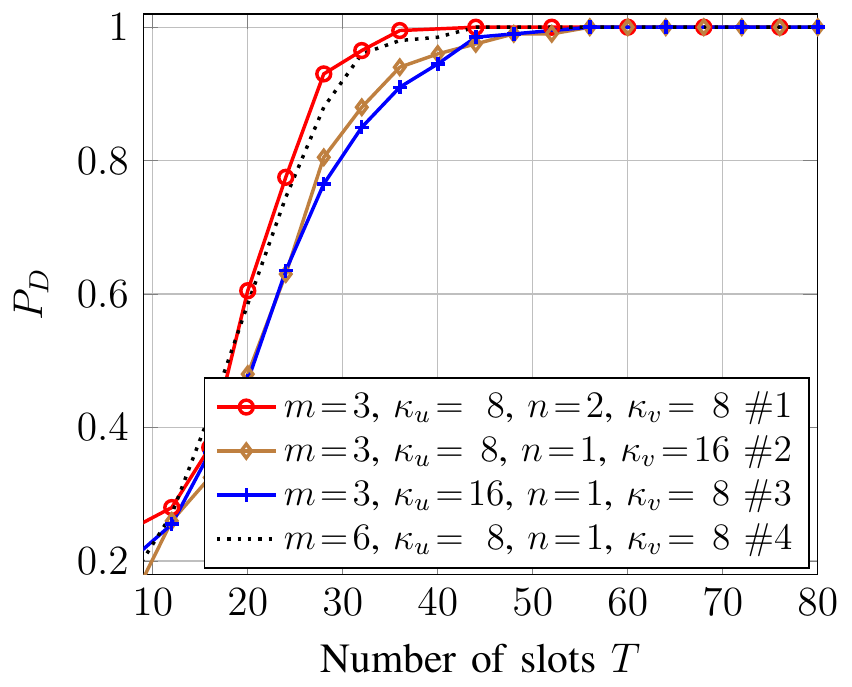}
	\caption{{\small {\em Detection probability $P_D$ of our proposed scheme when the product of $(mn\kappa_u\kappa_v)$, i.e., the probing dimensions in the spatial multiplexing domain, are constant, and where $M=N=32$, $m=3$, $n=2$, $F'=3$, $\snrbef=-33$ dB}}.}
	\label{Constant_mnkvku}
\end{figure}

{\bf Dependence on the number of RF chains $n$ for users.}
At the user side, increasing the number of RF chains provides more independent measurements, while on the other hand, splitting more of the received signal power. Obviously, when the channel pre-beamforming SNR is large enough, it is always beneficial to use more RF chains so as to get more independent measurements at one shot and to speed up the BA procedure. On the contrast, as shown in Fig.\,\ref{change_n}, when the pre-beamforming SNR is very low, e.g., $\snrbef=-36$ dB, it is better to just use one RF chain at the user side. For intermediate cases, however, there would be an optimal number of RF chains for the user, where fixing the BS and channel parameters, the BA procedure can be completed in the shortest time.  
\begin{figure}[t]
	\centering
	\includegraphics[width=8cm]{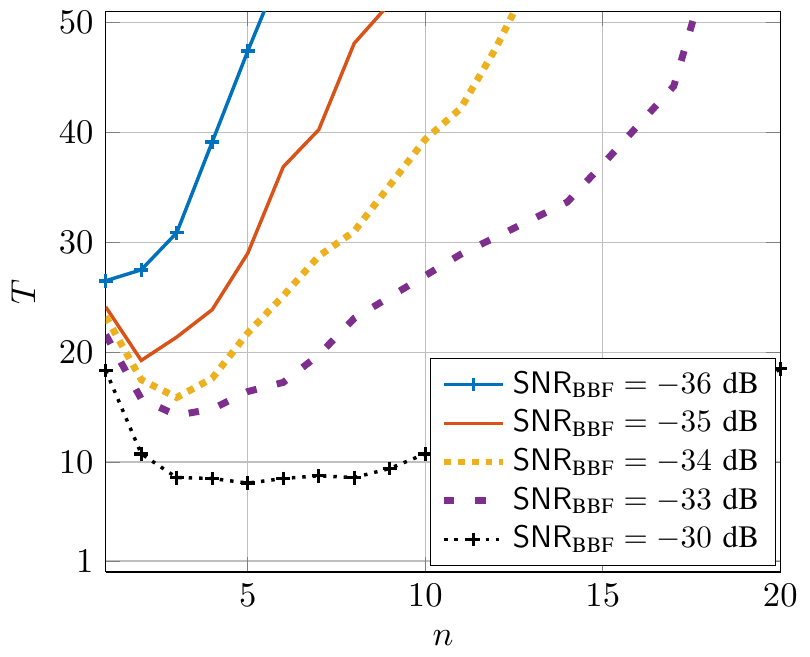}
	\caption{{\small {\em Expected detection time (slots) $T$ of our proposed scheme when increasing the number of RF chains  $n$ at the user side, where $M=N=32$, $F'=3$, $m=3$}}.}
	\label{change_n}
\end{figure}

{\bf System-Level Scalability.}
For a multi-user scenario, we denote by $K$ the total number of active users in the system, and by $K(T)$  the number of users that are able to successfully detect their strong MPC direction within $T$ frames. 
Fig.\,\ref{NNLSBisec} compares the fraction $\frac{K(T)}{K}$ of those users in our scheme with the corresponding fraction
in the interactive bisection method proposed in \cite{alkhateeb2014channel}, where we assume an ideal feedback and cost free for each iterative round in \cite{alkhateeb2014channel}. As we can see, the training overhead of interactive methods scales proportionally with the number of active users, 
whereas in our scheme all the users are essentially trained simultaneously, so that the overhead does not grow with the number of users. 
Note that in practice, the feedback scheme for each iterative round in \cite{alkhateeb2014channel} costs UL transmissions and may not be ideal 
since the beamforming gains are very poor at the initial rounds. In contrast, the proposed scheme needs only one UL transmission of the control packet, where the full beamforming gain at the UE side and the sectored beamforming gain at the BS side (as discussed in Section \ref{High-Level Overview}) are available. 

\begin{figure}[t]
	\centering
	\includegraphics[width=8cm]{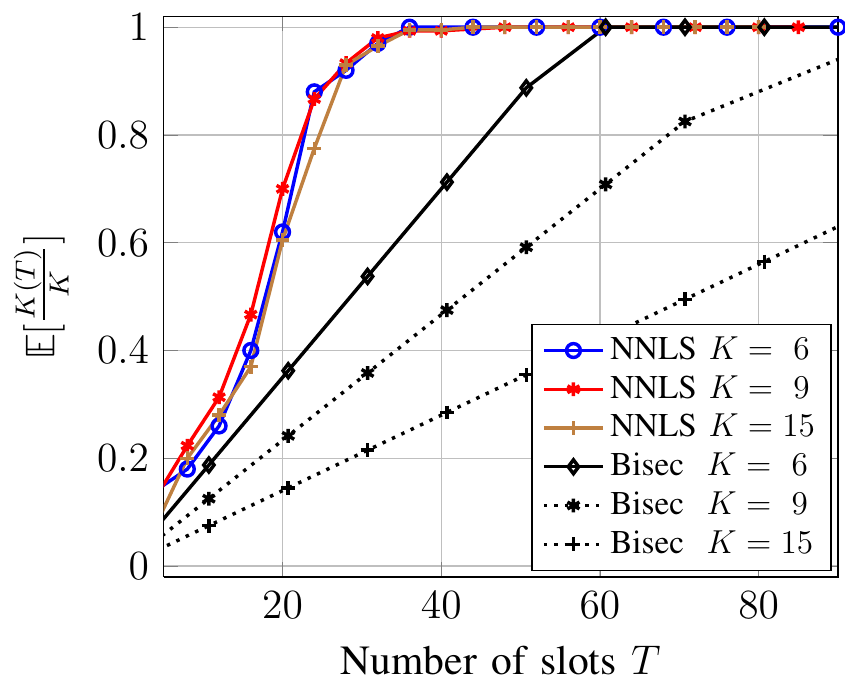}
	\caption{{\small {\em Comparison of the performance of our proposed scheme with that of interactive bisection method in \cite{alkhateeb2014channel} in terms of the fraction of users whose channel is estimated until a given time slot $T$ given by $\frac{K(T)}{K}$. We take $M=N=32$, $F'=3$, $m=3$, $n=2$, $\kappa_u=\kappa_v=8$, $\snrbef= -33$ dB}}.}
	\label{NNLSBisec}
\end{figure}

{\bf Robustness w.r.t. Variations in Channel Statistics.}
To investigate the sensitivity of the proposed scheme as well as competing CS-based schemes 
to the time-variation of the channel coefficients, we consider a simple Gauss-Markov model for the channel correlation in time given by 
\begin{align}  \label{bla}
\rho_{s,l}=\alpha {\rho}_{s-1,l}+\sqrt{1-|\alpha|^2}\, \nu_{s,l}, s\in \bZ_+,
\end{align}
where ${\rho_{0,l}\sim \cg(0, \gamma_l)}$, where $\nu_{s,l}\sim \cg(0, \gamma_l)$ is an i.i.d. sequence (innovation), and where 
$|\alpha|\in [0,1]$ controls the channel correlation in time. We assume that the channel is constant over each beacon slot of 14 OFDM symbols, 
and evolves in time according to (\ref{bla}) from slot to slot, i.e., $\alpha$ denotes the channel correlation coefficient at samples taken 1 ms apart.  
Fig.\,\ref{changeRHO} illustrates the comparison of the performance of our proposed scheme with that of 
the CS-based technique in \cite{ahmedmultiuser}. It is seen that our method exhibits much robust performance across a wide range of 
channel time-correlations whereas the algorithm in \cite{ahmedmultiuser} is quite fragile and fails to estimate the BA direction 
in the presence of channel time-variations from slot to slot. 

\begin{figure}[t]
	\centering
	\includegraphics[width=8cm]{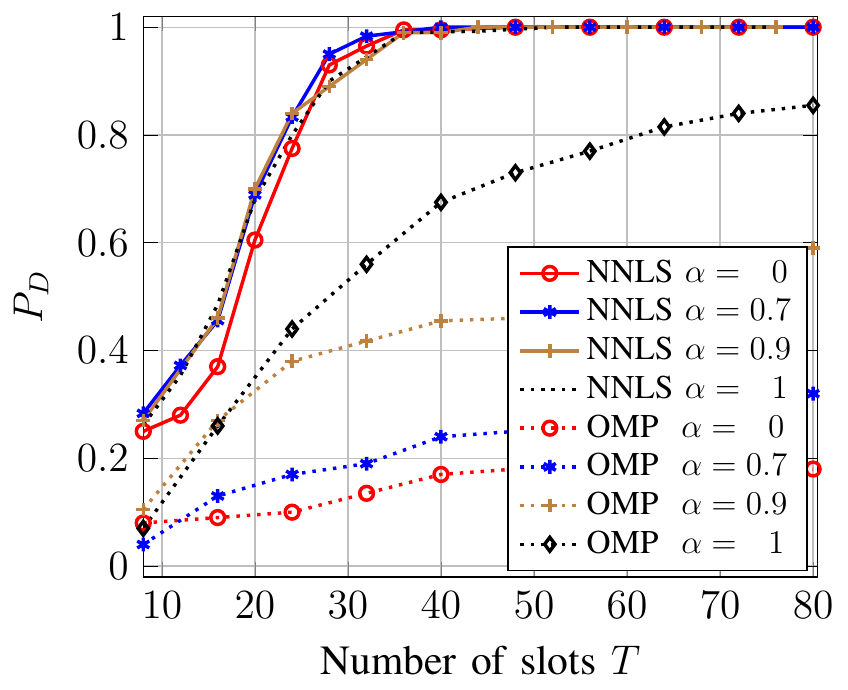}
	\caption{{\small {\em Comparison of detection probability $P_D$ between proposed NNLS and OMP in \cite{ahmedmultiuser} for different statistic path gains, where $M=N=32$, $F'=3$, $m=3$, $n=2$, $\kappa_u=\kappa_v=8$, $\snrbef= -33$ dB,  the path gains change from i.i.d. ($\alpha=0$) to constant ($\alpha=1$) over slots}}.}
	\label{changeRHO}
\end{figure}

\section{Conclusion}\label{conclusion}

In this paper, we proposed an efficient \textit{Beam Alignment} (BA) scheme for mm-Wave multiuser MIMO systems. 
In our proposed scheme, the channel is always probed by the BS in the \textit{Downlink} (DL), providing all the users within the BS coverage sufficiently 
many measurements to estimate the AoA-AoD of strong MPCs connecting them to the BS. In contrast with the conventional interactive BA algorithms, 
that require several rounds of beam refinement and transmissions of probing signals and/or control packets both in the DL and in the UL (\textit{Uplink}), in our proposed scheme all the users are trained simultaneously. Thus, the BA scales very well with the number of active users in the system. We posed the BA as the estimation of the second order statistics of the channel and proposed a novel technique based on NNLS that achieves reliable recovery with high probability.  We illustrated, via numerical simulations and comparison with other competitive techniques in the literature, that our algorithm is highly robust to variations in the 
channel statistics.

\section*{Acknowledgments}
X.S. is sponsored by the China Scholarship Council (201604910530). G.C. is supported by the 
Alexander von Humboldt Foundation through a Professorship Grant, and this work was also supported in part by 
a Collaborative Research Grant of Intel Research.

\balance
{\footnotesize
\bibliographystyle{IEEEtran}
\bibliography{references}
}

\end{document}

%% file: symbols.tex
\usepackage{mathbbol}

\def\mindex#1{\index{#1}}

\def\sq{\hbox{\rlap{$\sqcap$}$\sqcup$}}
\def\qed{\ifmmode\sq\else{\unskip\nobreak\hfil
\penalty50\hskip1em\null\nobreak\hfil\sq
\parfillskip=0pt\finalhyphendemerits=0\endgraf}\fi\medskip}

\long\def\defbox#1{\framebox[.9\hsize][c]{\parbox{.85\hsize}{%
\parindent=0pt
\baselineskip=12pt plus .1pt      
\parskip=6pt plus 1.5pt minus 1pt 
 #1}}}

\long\def\beginbox#1\endbox{\subsection*{}%
\hbox{\hspace{.05\hsize}\defbox{\medskip#1\bigskip}}%
\subsection*{}}

\def\endbox{}

\newsavebox{\junk}
\savebox{\junk}[1.6mm]{\hbox{$|\!|\!|$}}

\def\argmin{\mathop{\rm arg\, min}}

\newcommand{\field}[1]{\mathbb{#1}}

\def\Re{\field{R}}

\def\intgr{\field{Z}}

\def\bC{{\mathbb C}}

\def\bE{{\mathbb E}}

\def\bR{{\mathbb R}}

\def\bZ{{\mathbb Z}}

\def\bbh{{\mathbb h}}

\def\bfA{{\bf A}}
\def\bfB{{\bf B}}
\def\bfC{{\bf C}}

\def\bfF{{\bf F}}

\def\bfH{{\bf H}}

\def\bfa{{\bf a}}
\def\bfb{{\bf b}}

\def\bfg{{\bf g}}

\def\bfq{{\bf q}}
\def\bfr{{\bf r}}

\def\bfu{{\bf u}}
\def\bfv{{\bf v}}
\def\bfw{{\bf w}}
\def\bfx{{\bf x}}

\def\ttd{{\mathtt d}}

\def\tti{{\mathtt i}}

\def\ttm{{\mathtt m}}

\def\tto{{\mathtt o}}

\def\ttt{{\mathtt t}}

\def\sfH{{\sf H}}

\def\bfmath#1{{\mathchoice{\mbox{\boldmath$#1$}}%
{\mbox{\boldmath$#1$}}%
{\mbox{\boldmath$\scriptstyle#1$}}%
{\mbox{\boldmath$\scriptscriptstyle#1$}}}}

\def\bfmY{\bfmath{Y}}

\def\bfmhhaY{\bfmath{\hhaY}} 
\def\bfmhhaY{\hbox to 0pt{$\widehat{\bfmY}$\hss}\widehat{\phantom{\raise 1.25pt\hbox{$\bfmY$}}}}

\def\til={{\widetilde =}}

\def\clA{{\cal A}}
\def\clB{{\cal B}}
\def\clC{{\cal C}}

\def\clF{{\cal F}}
\def\clG{{\cal G}}

\def\clN{{\cal N}}

\def\clU{{\cal U}}
\def\clV{{\cal V}}

\def\clX{{\cal X}}

 \def\FRAC#1#2#3{\genfrac{}{}{}{#1}{#2}{#3}}

\def\ddtp{{\mathchoice{\FRAC{1}{d^{\hbox to 2pt{\rm\tiny +\hss}}}{dt}}%
{\FRAC{1}{d^{\hbox to 2pt{\rm\tiny +\hss}}}{dt}}%
{\FRAC{3}{d^{\hbox to 2pt{\rm\tiny +\hss}}}{dt}}%
{\FRAC{3}{d^{\hbox to 2pt{\rm\tiny +\hss}}}{dt}}}}

\def\average#1,#2,{{1\over #2} \sum_{#1}^{#2}}

\def\eye(#1){{\bf(#1)}\quad}

\newtheorem{remark}{{\bf Remark}}

\def\eq#1/{(\ref{e:#1})}

\newcommand{\inp}[2]{{\langle #1, #2 \rangle}}

\newcommand{\beqn}[1]{\notes{#1}%
\begin{eqnarray} \elabel{#1}}

\newcommand{\eeqn}{\end{eqnarray} }

\newcommand{\beq}[1]{\notes{#1}%
\begin{equation}\elabel{#1}}

\newcommand{\eeq}{\end{equation}}

\def\bdes{\begin{description}}
\def\edes{\end{description}}

\newcounter{rmnum}

\newcounter{anum}

{\end{list}}

\def\ass(#1:#2){(#1\ref{#1:#2})}

\def\ritem#1{
\item[{\sf \ass(\current_model:#1)}]
}

\newenvironment{recall-ass}[1]{%
\begin{description}
\def\current_model{#1}}{
\end{description}
}



%% file: macros.tex
\long\def\comment#1{}

\newcommand{\gv}{{\bf g}}

\newcommand{\uv}{{\bf u}}

\newcommand{\vv}{{\bf v}}

\newcommand{\Bm}{{\bf B}}

\newcommand{\Fm}{{\bf F}}

\newcommand{\Hm}{{\bf H}}

\newcommand{\Cc}{{\cal C}}

\newcommand{\Fc}{{\cal F}}

\newcommand{\Uc}{{\cal U}}

\newcommand{\Vc}{{\cal V}}

\newcommand{\gammav}{\hbox{\boldmath$\gamma$}}

\newcommand{\Gammam}{\boldsymbol{\Gamma}}

\newcommand{\gammam}{\boldsymbol{\gamma}}

\newcommand{\Sigmam}{\hbox{\boldmath$\Sigma$}}

\renewcommand{\Re}{{\rm Re}}

\newcommand{\transp}{{\sf T}}
\renewcommand{\vec}{{\rm vec}}

